\documentclass[11pt]{article}
\usepackage[letterpaper,margin=1in]{geometry}
\usepackage[T1]{fontenc}
\usepackage{lmodern,microtype,amsmath,amssymb,amsthm,mathtools,aliascnt,booktabs}
\usepackage[dvipsnames]{xcolor}
\definecolor{linkblue}{RGB}{25,75,130}
\usepackage[colorlinks=true,linkcolor=linkblue,urlcolor=linkblue,citecolor=linkblue]{hyperref}
\usepackage{enumitem,float,needspace}
\usepackage{algorithm,algpseudocode}
\algrenewcommand\algorithmicrequire{\textbf{Input:}}
\algrenewcommand\algorithmicensure{\textbf{Output:}}
\usepackage{graphicx}
\hypersetup{pdftitle={Breaking the quantum barriers for Graph Collision},pdfauthor={Arjan Cornelissen; Amin Shiraz Gilani; Shiva Kintali; Fran\c{c}ois Le Gall; Subhasree Patro; Xingyu Zhou}}
\allowdisplaybreaks
\setlist{itemsep=0.25em,topsep=0.5em}
\usepackage{bbm}

\usepackage{soul}
\usepackage[nameinlink,noabbrev,capitalise]{cleveref}
\newtheorem{theorem}{Theorem}[section]
\newaliascnt{lemma}{theorem}
\newtheorem{lemma}[lemma]{Lemma}
\aliascntresetthe{lemma}
\newcommand{\restatedlemmaname}{}
\newtheorem*{repeatedlemma}{\restatedlemmaname}
\newenvironment{restatedlemma}[1]{\renewcommand{\restatedlemmaname}{Lemma~\ref{#1} (restated)}\begin{repeatedlemma}}{\end{repeatedlemma}}
\newaliascnt{corollary}{theorem}
\newtheorem{corollary}[corollary]{Corollary}
\aliascntresetthe{corollary}
\theoremstyle{definition}
\newaliascnt{definition}{theorem}
\newtheorem{definition}[definition]{Definition}
\aliascntresetthe{definition}

\crefname{theorem}{Theorem}{Theorems}
\crefname{lemma}{Lemma}{Lemmas}
\crefname{corollary}{Corollary}{Corollaries}
\crefname{definition}{Definition}{Definitions}

\numberwithin{equation}{section}
\AddToHook{cmd/thebibliography/after}{\setlength{\itemsep}{0pt}}

\newenvironment{steps}{\begin{enumerate}[label=\arabic*:,font=\footnotesize,leftmargin=1.9em,labelsep=.5em,itemsep=1pt,topsep=3pt,parsep=0pt]}{\end{enumerate}}
\newcommand{\ind}{\hspace*{1em}}
\newcommand{\Yes}{\textnormal{\textsc{Yes}}}
\newcommand{\No}{\textnormal{\textsc{No}}}
\newcommand{\Ot}{\widetilde O}
\newcommand{\vol}{\operatorname{vol}}
\newcommand{\sampleneighborhoods}{\ensuremath{\mathsf{sample\_neighborhoods}}}
\newcommand{\covertest}{\ensuremath{\mathsf{cover\_test}}}
\newcommand{\bipartitetest}{\ensuremath{\mathsf{bipartite\_test}}}
\newcommand{\capdegrees}{\ensuremath{\mathsf{cap\_degrees}}}
\newcommand{\partitionedges}{\ensuremath{\mathsf{partition\_edges}}}
\newcommand{\prepareestimates}{\ensuremath{\mathsf{prepare\_estimates}}}
\newcommand{\estimatesums}{\ensuremath{\mathsf{estimate\_sums}}}
\newcommand{\reducevolume}{\ensuremath{\mathsf{reduce\_volume}}}
\newcommand{\extractneighborhoodedges}{\ensuremath{\mathsf{extract\_neighborhood\_edges}}}
\newcommand{\estimateneighborhoodcounts}{\ensuremath{\mathsf{estimate\_neighborhood\_counts}}}
\newcommand{\samplemarked}{\ensuremath{\mathsf{sample\_marked\_neighborhoods}}}

\title{Breaking the quantum barriers for Graph Collision}
\author{Arjan Cornelissen\thanks{University of Cambridge.
    \href{mailto:ajcornelissen@outlook.com}{\texttt{ajcornelissen@outlook.com}}}
  \qquad
  Amin Shiraz Gilani\thanks{University of Maryland.
    \href{mailto:asgilani@umd.edu}{\texttt{asgilani@umd.edu}}}
  \qquad
  Shiva Kintali\thanks{NVIDIA.
    \href{mailto:shiva.kintali@gmail.com}{\texttt{shiva.kintali@gmail.com}}}\\[0.6em]
  Fran\c{c}ois Le Gall\thanks{Nagoya University.
    \href{mailto:legall@math.nagoya-u.ac.jp}{\texttt{legall@math.nagoya-u.ac.jp}}}
  \qquad
  Subhasree Patro\thanks{Eindhoven University of Technology.
    \href{mailto:s.patro@tue.nl}{\texttt{s.patro@tue.nl}}}
  \qquad
  Xingyu Zhou\thanks{The University of British Columbia.
    \href{mailto:zxingyu@cs.ubc.ca}{\texttt{zxingyu@cs.ubc.ca}}}}
\date{October 2026}
\begin{document}
	\maketitle

	\begin{abstract}
		In the graph collision problem, the task is to decide whether a known graph on $n$ vertices contains an edge with both endpoints marked, given query access to the input bits specifying the marked vertices. We first refute the conjectured $\Omega(n^{2/3})$ query lower bound with a bounded error quantum algorithm using $\widetilde O(n^{92/139})$ queries. We then refine this approach to obtain a $\widetilde O(n^{11/17})$ query algorithm. We also obtain a $\widetilde O((nq)^{11/17})$ query bound for all (partial) Boolean output functions with positive certificate complexity at most 2 and alphabet size $q$, contrasting a known $\Omega(n^{2/3})$ lower bound for such problems with large alphabet sizes. We complement these algorithms with an $\Omega(\sqrt{n\log n})$ quantum query lower bound for graph collision, exceeding the $O(\sqrt n)$ certificate complexity barrier for the positive adversary method.
	\end{abstract}

	\clearpage
	\begingroup
	\setcounter{tocdepth}{2}
	\tableofcontents
	\endgroup
	\clearpage
	
	\section{Introduction}

In the graph collision problem, a simple undirected graph $G=(V,E)$ on $n$ vertices is known in advance, and we have query access to an input $x\in\{0,1\}^V$. A vertex $v$ is marked if $x_v=1$. The task is to decide whether some edge $uv\in E$ has both endpoints marked. Only queries to $x$ are counted; all computation involving the known graph is free.

Magniez, Santha, and Szegedy introduced graph collision as a subroutine for triangle finding and gave an $O(n^{2/3})$ query quantum algorithm for arbitrary graphs~\cite{MSS,jeffery2016improving}. Their algorithm uses Ambainis's quantum walk for element distinctness~\cite{ambainis2007quantum}. Subsequent work improved this bound for graphs with few nonedges~\cite{jeffery2016improving}, small independence number~\cite{belovs2012learning}, or small treewidth~\cite{AB}. Other bounds depend on the degrees of vertices in independent sets~\cite{GI,AB}. Despite these improvements for particular graph families, the $O(n^{2/3})$ bound remained the best known for arbitrary graphs.

A reduction from unstructured search gives a worst case lower bound of $\Omega(\sqrt n)$~\cite{buhrman1999bounds}, leaving a gap between the known upper and lower bounds. Gavinsky and Ito ask whether graph collision admits a $\widetilde O(\sqrt n)$ query algorithm~\cite[Section~6]{GI}, while Ambainis et al.\ and Balodis and Iraids leave an $O(\sqrt n)$ bound for arbitrary graphs open~\cite{AB,BI16}. Determining the quantum query complexity of graph collision also appears among the open problems in Kothari's thesis~\cite[Open Problem~4.3]{Kot14} and in the work of Bun, Kothari, and Thaler~\cite[Section~7.2]{BKT20}. In the other direction, Cornelissen, Gilani, and Patro report that the $n^{2/3}$ bound is believed by some in the community to be optimal~\cite{CGP26}. Gilani, Mande, Patro, and Raja formulate a conjecture for detecting marked cliques whose $k=2$ case asserts that some graph families require $\Omega(n^{2/3})$ queries for graph collision~\cite[Conjecture~1.5]{GMPR26}.

We refute this conjecture. We first give a relatively elementary algorithm using $\Ot(n^{92/139})$ queries in Section~\ref{sec:simple}. Its central idea is to estimate the number of marked neighbors of each vertex and use these estimates to divide the edges into parts that can be tested efficiently. In Section~\ref{sec:algorithm}, we refine the same approach to obtain the following stronger bound.

\begin{theorem}\label{thm:main}
    There is a quantum algorithm that solves the graph collision problem on any $n$-vertex graph with high probability using $\Ot(n^{11/17})$ queries.
\end{theorem}

Our upper bound also applies to functions whose positive inputs can be certified by fixing at most 2 input symbols. In Section~\ref{sec:certificates}, we reduce any such (partial) Boolean output function on $n$ input symbols from an alphabet of size $q$ to graph collision on $nq$ vertices. By \Cref{thm:main}, this gives a query bound of $\widetilde O((nq)^{11/17})$. This contrasts with the $\Omega(n^{2/3})$ lower bound for element distinctness over large alphabets~\cite{AS04,ambainis2005smallrange,rosmanis2014quantum}, whose positive inputs also have certificates of size 2. More generally, our bound is $o(n^{2/3})$ for $q=O(n^{1/33-\varepsilon})$ and fixed $\varepsilon>0$, showing the importance of the alphabet size in lower bounds based on certificate structures~\cite{belovs2014power}.

We also improve the lower bound for graph collision. For a fixed graph this is a total Boolean function with positive certificates of size 2, so the positive adversary method cannot prove a lower bound larger than $O(\sqrt n)$~\cite{Zha05,SS06}. We overcome this certificate barrier using the general adversary method, which allows negative weights~\cite{HLS07}. Our explicit adversary matrix construction is inspired by the recent framework of Belovs~\cite{belovs2026tight} and gives the following result.

\begin{theorem}[Informal version of \Cref{thm:incidence-lower-bound}]
    There is a family of graphs on $n$ vertices for which the quantum query complexity of graph collision is $\Omega(\sqrt{n\log n})$.
\end{theorem}

This rules out an $O(\sqrt n)$ query algorithm for arbitrary graphs, while leaving open the possibility of a $\widetilde O(\sqrt n)$ query algorithm.
	
	\subsection{Technical overview}

    We give an informal overview of the main ideas behind our two algorithms, the reduction for functions with small positive certificates, and the lower bound. Precise statements and proofs appear in Sections~\ref{sec:simple}--\ref{sec:lower-bound}.

	\subsubsection{An algorithm breaking the \texorpdfstring{$O(n^{2/3})$}{O(n\textasciicircum(2/3))} barrier}
	
	We describe the ideas behind the algorithm of Section~\ref{sec:simple}, which achieves query complexity $\Ot(n^{92/139})$. The algorithm first preprocesses the graph, then estimates neighborhood counts and uses a threshold to divide the remaining edges into two parts. We begin by explaining why such estimates suffice. The improved algorithm of Section~\ref{sec:algorithm} builds on the same estimation and thresholding argument.
	
	Let $G=(V,E)$ be the known graph, and let $M=\{v:x_v=1\}$ be its unknown marked set. For any set $S\subseteq V$, write $\ell_u(S)=|N_G(u)\cap S|$, where $N_G(u)$ is the neighborhood of $u$ in the graph $G$. In particular, $\ell_u(M)$ counts the marked neighbors of $u$. Suppose for a moment that we knew all these counts. We could then restrict attention to vertices with positive counts and use Grover search to find a marked vertex among them. This would decide graph collision using $O(\sqrt n)$ additional queries.
	
	\paragraph{Approximate counts and marked volume.}
	Computing all the counts exactly would require too many queries. We instead construct a set $S$ for which the counts $\ell_u(S)$ approximate $\ell_u(M)$. Our aim is to bound their total squared error
	\begin{equation}
        \label{eq:total-squared-error}
	    r=\sum_u\bigl(\ell_u(S)-\ell_u(M)\bigr)^2.
	\end{equation}
	The set $S$ need not recover the marked set. We only use its neighborhood counts. Although approximate counts need not identify which vertices have a marked neighbor, they suffice to divide the edges into two parts that can be tested efficiently.
	
	The relevant quantity for these tests is the \emph{marked volume} $\vol(G,x)=\sum_{v\in M}\deg_G(v)$, the total degree of marked vertices. Given an upper bound $b$ on this quantity, variable-time search gives a collision test using $\Ot(\sqrt n+\sqrt b)$ queries.
	
	Choose a threshold $\tau>0$ and let $P=\{u:\ell_u(S)\le\tau\}$. Consider first the edges inside $P$. The marked volume of $G[P]$ equals $\sum_{u\in P}|N_G(u)\cap M\cap P|$. Each term in this sum is at most $\ell_u(M)$, which also counts marked neighbors outside $P$. The estimated counts are at most $\tau$ on $P$, and Cauchy--Schwarz bounds the sum of the absolute errors by $\sqrt{nr}$. Consequently,
	\[
	 \vol(G[P],x)\le\sum_{u\in P}\ell_u(M)
	 \le n\tau+\sqrt{nr}.
	\]
	For the edges with at least one endpoint outside $P$, the set $V\setminus P$ contains an endpoint of every such edge and is therefore a vertex cover. The same test applies with a bound on $\vol(G,V\setminus P,x)=\sum_{v\in M\setminus P}\deg_G(v)$, the sum of the degrees of marked vertices in this cover. On an input with no collision, every marked vertex has zero marked neighbors. A marked vertex outside $P$ must therefore contribute more than $\tau^2$ to the squared error, which gives $|M\setminus P|\le r/\tau^2$. If the maximum degree is at most $d$, the marked volume of this cover is at most $dr/\tau^2$.

	Thus the same error bound controls both edge sets: the counts inside $P$ bound their marked volume, while inaccurate counts at marked vertices outside $P$ limit the marked volume of the other cover. We apply the same volume test to both parts, using $P$ as the cover inside and $V\setminus P$ as the cover outside. The volume assumptions are needed for rejection on negative inputs; a test detects a collision in its edge set with high probability even if the stated volume bound fails. We can therefore analyze the estimation error on negative inputs and still detect collisions on positive inputs.

\begin{figure}[H]
		\centering
		\includegraphics[width=.94\linewidth]{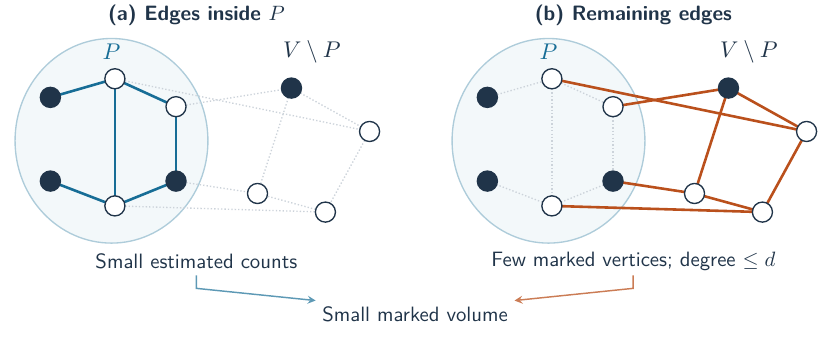}
		\caption{Thresholding the estimated marked neighbor counts partitions the edges. Inside $P$, small estimates and controlled error bound the marked volume of the blue graph. On a negative input, every marked vertex outside $P$ has a large estimation error, so there are few such vertices. Their degrees are at most $d$, giving a small marked volume for the outside cover of the orange edges. Filled vertices are marked; dotted edges belong to the other part.}
		\label{fig:thresholding}
	\end{figure}

	Figure~\ref{fig:thresholding} illustrates the two marked volume bounds.

	\paragraph{Estimating neighborhood counts.}
	The main question is how to obtain the approximate counts with few queries. Take $q$ independent uniform vertices $w_1,\ldots,w_q$, query their input bits, and form the estimates
	\[
	 y_u=\frac nq\sum_{i=1}^q x_{w_i}\mathbf1[w_i\in N_G(u)].
	\]
	Each estimate is unbiased, but its sampling error may be large. The true neighborhood counts all come from the same marked set $M$. We use this common structure by choosing a single set $S\subseteq V$ minimizing
	\[
	 \sum_u\bigl(\ell_u(S)-y_u\bigr)^2.
	\]
	The crucial observation is that the true set $M$ is an allowed choice, so the objective value at $S$ is no larger than at $M$. We can use this inequality to compare the neighborhood counts of $S$ and $M$, even though we do not know $M$.
	
	To analyze this comparison, let $c_{vw}=|N_G(v)\cap N_G(w)|$ be the number of common neighbors of $v$ and $w$. Define $z_v=\sum_{u\in N_G(v)}y_u$, which estimates the number of walks of length two from $v$ to marked vertices. A sampled marked vertex $w_i$ contributes $n/q$ to this sum once for each common neighbor of $v$ and $w_i$, so
	\[
	 z_v=\frac nq\sum_{i=1}^q c_{vw_i}x_{w_i},
	 \qquad \mathbb E[z_v]=\sum_{w\in M}c_{vw}.
	\]
	Counting in two ways gives $\sum_u\ell_u(S)y_u=\sum_{v\in S}z_v$. Expanding the least squares objective leaves $\psi_z(S)=\sum_u\ell_u(S)^2-2\sum_{v\in S}z_v$, together with a term $\sum_u y_u^2$ that is independent of $S$. Thus it suffices to compute the common neighbor estimates $z_v$ and minimize $\psi_z$. All these computations use the known graph and the sampled input bits, so they require no further queries. Algorithm~\ref{alg:fit} gives the procedure.

	Comparing $\psi_z(S)$ with $\psi_z(M)$ now bounds the squared estimation error by
	\[
	 r\le2\sum_v\left|z_v-\mathbb E[z_v]\right|.
	\]
	This is the useful consequence of choosing all the estimated counts from one set: their total squared error is controlled by the sum of the absolute errors in the sampled common neighbor sums. The improved algorithm uses the same comparison.
	
	The variance of $z_v$ is at most $(n/q)\sum_{w\in M}c_{vw}^2$. Applying Cauchy--Schwarz to the comparison above yields
	\[
	 \mathbb E[r]\le
	 2\sqrt{n\sum_v\operatorname{Var}(z_v)}
	 \le\frac{2n}{\sqrt q}
	 \left(\sum_{w\in M}\sum_v c_{vw}^2\right)^{1/2}.
	\]
	For fixed $v,w$, the quantity $c_{vw}^2$ counts ordered pairs of common neighbors, or equivalently closed walks $w,a,v,b,w$ of length four. Only walks starting at marked vertices contribute, allowing us to use the reduction in marked volume obtained by sampling even if many edges remain.
	
	\paragraph{Sampling and edge deletion.}
	The squared common neighbor counts may be too large in an arbitrary graph. We therefore preprocess the graph before estimating the counts. A separate, earlier round of sampling reduces the marked volume; further edge deletions bound the degrees and the number of closed walks starting at marked vertices. Each deletion is accompanied by a search or collision test that detects a collision among the deleted edges with high probability. The algorithm returns as soon as a test finds one.
	
	First, take $\Theta(q\log n)$ uniform vertex samples and query their input bits. Search the union of the neighborhoods of the marked samples for a marked vertex. A successful search gives a collision. Otherwise, delete all edges incident to this union. Any vertex with more than $n/q$ marked neighbors is likely to have one of those neighbors sampled, in which case the deletion isolates it. With high probability every remaining vertex has at most $n/q$ marked neighbors. Summing these counts gives $\vol(G,x)\le n^2/q$. Algorithm~\ref{alg:sample} gives this sampling procedure.
	
	Next, consider the vertices of degree greater than $d$. Each marked vertex among them contributes more than $d$ to the marked volume, so there are at most $n^2/(qd)$ such marked vertices. The test of Lemma~\ref{lem:few} uses this count bound to check all their incident edges. If the test rejects, delete those edges. The remaining graph has maximum degree at most $d$, and its marked volume is still at most $n^2/q$. The degree bound has two uses: it converts the bound on $|M\setminus P|$ into a marked volume bound, and it makes the neighborhoods in the next procedure small.
	
	Finally, whenever an edge $uv$ in the current graph has at least $h$ edges between $N_G(u)$ and $N_G(v)$, record and delete that edge set. Repeat until no such edge remains. There are at most $nd/h$ recorded graphs, since their edge sets are disjoint and each contains at least $h$ edges. We use Grover search over these graphs to find one containing a collision. Each graph has at most $2d$ vertices and can be tested using a standard quantum walk. Algorithm~\ref{alg:basic} gives the extraction and search procedure.

\begin{figure}[H]
		\centering
		\includegraphics[width=.94\linewidth]{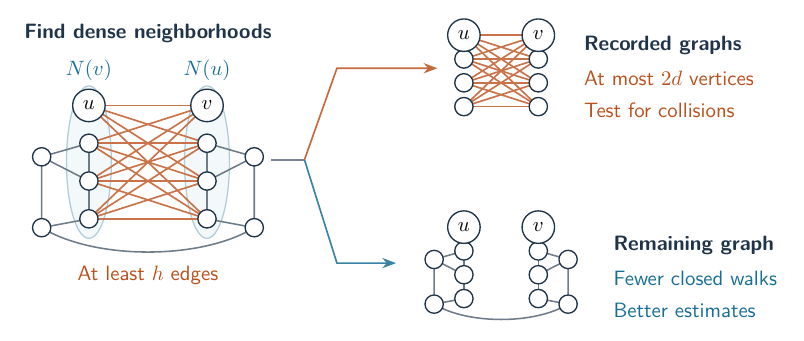}
		\caption{When an edge $uv$ has at least $h$ edges between $N(u)$ and $N(v)$, extract those edges, shown in orange. The upper branch tests the extracted graph for a collision; it has at most $2d$ vertices because the maximum degree is at most $d$. Repeating the extraction leaves the graph in the lower branch with fewer closed walks of length four starting at marked vertices. This controls the error in the neighborhood estimates.}
		\label{fig:extraction}
	\end{figure}

	Figure~\ref{fig:extraction} summarizes how extraction supports both collision testing and estimation.

	To see why this deletion controls the required walks, fix the first directed edge $wa$ of a closed walk $w,a,v,b,w$. The edge $vb$ joins $N_G(a)$ to $N_G(w)$, and choosing its orientation determines the whole walk. At termination there are fewer than $h$ choices for this edge, with at most two possible orientations. There are $\vol(G,x)$ choices of the first directed edge with a marked initial vertex. Hence the remaining graph satisfies
	\[
	 \sum_{w\in M}\sum_v c_{vw}^2\le2h\,\vol(G,x).
	\]
	A larger $h$ reduces the number of recorded graphs but allows a larger estimation error. Taking fresh samples for estimation after preprocessing, and using $\vol(G,x)\le n^2/q$, gives $\mathbb E[r]=O(n^2\sqrt h/q)$. Markov's inequality makes a sufficiently large constant multiple of this quantity a valid error bound with high constant probability.
	
	Algorithm~\ref{alg:simple} combines these procedures in their execution order: sample and delete edges, test the high degree vertices, extract and test the graphs between neighborhoods, take fresh samples for estimation, and apply the threshold and the two final volume tests. A collision is either detected among the deleted edges or remains in one of the final edge sets. Balancing the sampling and preprocessing costs with the two final volume bounds gives query complexity $\Ot(n^{92/139})$. Section~\ref{sec:simple} chooses the parameters after deriving these costs.
	
	\subsubsection{An improved algorithm}

    As in Algorithm~\ref{alg:fit}, we estimate the common neighbor sums $\sum_{w\in M}c_{vw}$ by $z_v$ and choose a set $S$ minimizing
    \[
        \psi_z(S)=\sum_u\ell_u(S)^2-2\sum_{v\in S}z_v.
    \]
    The resulting counts $\ell_u(S)=|N_G(u)\cap S|$ approximate the actual marked neighbor counts. Thresholding these counts determines the set $P$, as in the first algorithm. The main refinement is to use a bound on the marked volume to restrict the sets $S$ considered in the minimization. We first reduce the maximum degree, then use this stronger estimation argument to reduce the marked volume repeatedly.

    \paragraph{Degree reduction.}
    Choose a degree bound $d$, and let $H$ contain all vertices whose current degree exceeds $d$. For intuition, suppose we know $k=|M\cap H|$. When $k>(n/d)^{3/2}$, take $\Theta((n/d)\log n)$ uniform samples from the marked vertices in $H$ and search the union of their neighborhoods for a marked vertex. A successful search gives a collision. Otherwise, delete all edges incident to this union. This is the search and deletion used in Algorithm~\ref{alg:sample}, now sampling from the marked vertices in $H$.

    A vertex with many marked neighbors in $H$ is likely to have one of them sampled, so the deletion isolates it. With high probability, every remaining vertex has at most $kd/(2n)$ marked neighbors in the old set $H$. Counting in two ways bounds the sum of the remaining degrees of these marked vertices by $kd/2$. Hence at most $k/2$ marked vertices still have degree greater than $d$. We update $H$ and repeat.

    After logarithmically many repetitions, $H$ contains at most $(n/d)^{3/2}$ marked vertices. Although $H$ itself may be large, this bound lets us use Lemma~\ref{lem:few} to test all edges with at least one endpoint in $H$. If no collision is found, we delete all these edges. The remaining graph has maximum degree at most $d$. Algorithm~\ref{alg:cap} gives this degree reduction procedure, with total query cost $\Ot(n^{3/4}/d^{1/4})$.

    \paragraph{Repeated volume reduction.}
    The degree bound gives marked volume at most $nd$. Let $\ell$ be its current upper bound. In the first algorithm, thresholding is followed by collision tests on both the edges inside $P$ and the edges outside it. Here, we test the edges outside $P$ and continue reducing the graph inside $P$. Our target in each round is only to halve the marked volume.

    For a general graph, we apply this argument to bipartite subgraphs separately. In each one, $P$ consists of vertices with small estimated marked neighbor counts. The same thresholding argument as in the first algorithm bounds the marked volume retained inside $P$ and that of the outside cover used to test the discarded edges. We choose the parameters so that, on a negative input, the total retained volume is at most $\ell/2$. We combine the retained edges and repeat until the volume is small enough for the final collision test. Algorithm~\ref{alg:bounded-degree} gives this iteration.

    \paragraph{Estimating neighborhood counts.}
    To see how the current volume bound helps, suppose for a moment that every vertex has degree $d$. Then $|M|\le\ell/d$, so we minimize $\psi_z(S)$ subject to $|S|\le\ell/d$. The true marked set remains an allowed choice. If every common neighbor sum is estimated within error $\eta$, the comparison $\psi_z(S)\le\psi_z(M)$ gives total squared error at most $4\eta\ell/d$. Only vertices in $S$ or $M$ contribute to this comparison, and both sets have size at most $\ell/d$. A smaller volume bound therefore limits how many estimation errors can contribute.

    For a general graph, we use a result of Chlamt\'{a}\v{c}, Dinitz and Robinson~\cite{CDR19} to partition its edges into a polylogarithmic number of bipartite subgraphs with approximately uniform degrees on each side. Lower bounds on the degrees now give bounds on the number of marked vertices on each side. We impose these bounds in the minimization and choose a separate threshold for each side. The partition uses only the known graph and requires no queries.

    Algorithm~\ref{alg:fit} forms its estimates from uniform vertex samples. Here, we sample marked vertices on the appropriate side and average their common neighbor counts, scaled by the estimated number of marked vertices. The same samples give estimates for every vertex, since the graph is known. When the sums $\sum_{v\in M}c_{uv}^2$ are small for every $u$, a variance bound shows that few samples give accurate estimates.

    To handle large common neighbor counts, we extend the extraction procedure of Algorithm~\ref{alg:basic}. That procedure extracts edges between two neighborhoods. Here, we use a neighborhood $N_G(u)$ and a set $R$ of vertices on the same side as $u$. The number of edges between them is $\sum_{v\in R}c_{uv}$, so large common neighbor counts identify graphs that we can test separately. We select such sets $R$, test their edges using quantum walks, and delete them. This makes the remaining sums easier to estimate while checking the discarded edges for collisions. Algorithms~\ref{alg:extract} and~\ref{alg:estimate} give these procedures.

    The parameter analysis in \Cref{sec:algorithm} balances the degree reduction with the repeated volume reduction. Choosing $d=\Theta(n^{7/17})$ gives the query bound $\Ot(n^{11/17})$.
	
	\subsubsection{Functions with positive certificates of size two}
	Section~\ref{sec:certificates} shows that graph collision also captures the task of evaluating a (partial) function when every positive input has a certificate of size (at most) two. The idea is to represent possible input assignments by vertices and positive certificates by edges.
	
	Suppose the input has $n$ symbols from an alphabet of size $q$. Create a vertex $(i,a)$ for each position $i$ and alphabet symbol $a$, and mark it exactly when $x_i=a$. There are $nq$ vertices, with exactly one marked vertex for each input position. Whenever $x_i=a$ and $x_j=b$ form a positive certificate, add an edge between $(i,a)$ and $(j,b)$. Its endpoints are both marked exactly when the input satisfies that certificate. It follows that the constructed graph has a marked edge if and only if the original input is positive.
	
	A query at vertex $(i,a)$ only asks whether $x_i=a$, so it can be simulated with a constant number of queries to the original input. Applying the graph collision algorithm therefore gives a query bound of $\Ot((nq)^{11/17})$. The dependence on the alphabet comes from the increase in graph size. For constant alphabets the bound is $\Ot(n^{11/17})$, and for $q=n^{o(1)}$ it is $n^{11/17+o(1)}$. More generally, the complexity remains $o(n^{2/3})$ whenever $q=O(n^{1/33-\varepsilon})$ for a fixed $\varepsilon>0$.
	
	For binary inputs, the graph has $2n$ vertices and each vertex query uses one input query. Conversely, every graph collision problem has positive certificates of size at most two. Thus graph collision is as hard as any Boolean function on binary inputs with positive certificate complexity at most two, up to a constant factor in the input size (Corollary~\ref{cor:certificate-equivalence}).
	
	\subsubsection{A lower bound breaking the \texorpdfstring{$\Omega(\sqrt n)$}{Omega(sqrt(n))} barrier}
	Section~\ref{sec:lower-bound} proves an $\Omega(\sqrt{n\log n})$ lower bound. For a prime power $q$, let $G_q$ be the bipartite graph with one vertex for each point of $\mathbb F_q^2$ and one vertex for each nonvertical line. A line with slope $s$ and intercept $t$ consists of the points $(u,v)$ satisfying $v=su+t$ in $\mathbb F_q$. A point vertex and a line vertex are adjacent when this equality holds. There are $q^2$ vertices on each side, every vertex has degree $q$, and the graph has $q^3$ edges. The geometric property we use is that two distinct points lie on at most one common line, and two distinct lines meet in at most one point.

	We compare inputs with many marked points and lines but no collision to inputs with exactly one marked edge. In our construction, a positive input is obtained from a negative input by changing bits only at one line and its incident points.

	\paragraph{Choosing the inputs.}
	A negative input is sampled in two stages:
	\begin{enumerate}
		\item Mark each line independently with probability $r$.
		\item Mark each point lying on no marked line independently with probability $p$. Leave every other point unmarked.
	\end{enumerate}
	Call a point \emph{uncovered} if it lies on no marked line. The sampling rule prevents every collision, since a marked line forces all its incident points to be unmarked. Each point is uncovered with probability $(1-r)^q$, so its probability of being marked is $p(1-r)^q$.

	We want many marked vertices on both sides. Marking more lines, however, leaves fewer uncovered points. We compensate by increasing $p$, choosing $p(1-r)^{q-1}=r$. Then a point is marked with probability $r(1-r)$, almost the same as the line probability $r$. We take $r$ of order $\ln q/q$, with a sufficiently small constant factor. Each side then has order $q\log q$ marked vertices in expectation, while $p$ still tends to zero. This is the largest possible order of $r$ under the balancing condition: if $r$ grew asymptotically faster than $\ln q/q$, the probability of being uncovered would be too small to compensate with any $p\le1$.

\begin{figure}[!ht]
		\centering
		\includegraphics[width=\linewidth]{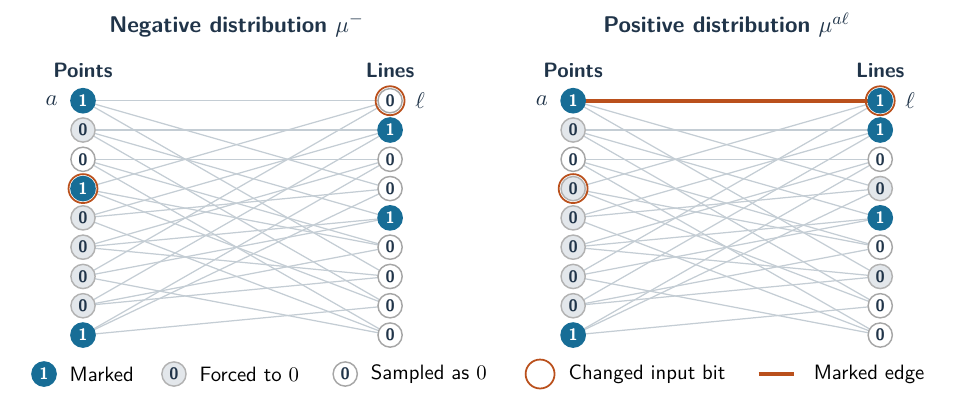}
		\caption{The hard input distributions, illustrated on $G_3$. On the left, lines are marked independently, and only uncovered points may be marked. Condition on $a$ being marked. On the right, marking $\ell$ and unmarking its other points creates the unique marked edge $a\ell$, shown in orange. Blue vertices are marked, gray vertices are forced to zero, and white vertices are sampled as zero. Orange rings identify exactly the input bits that change between the panels.}
		\label{fig:hard-distributions}
	\end{figure}

	To sample a positive input:
\begin{enumerate}
    \item Choose a point $a$ uniformly at random, then choose a line $\ell$ through $a$ uniformly at random. This chooses an edge $a\ell$ uniformly at random.
    \item Sample a negative input conditioned on $a$ being marked.
    \item Mark $\ell$ and unmark every point on $\ell$ other than $a$.
\end{enumerate}
All lines through $a$ were initially unmarked, so the resulting input has exactly one marked edge, namely $a\ell$. We call the last step \emph{planting} the edge $a\ell$. Once the edge and negative input are fixed, this operation determines a unique positive input, although different negative inputs can produce the same positive input. Each fixed edge defines its own positive distribution. These distributions have disjoint supports because the unique marked edge identifies the chosen edge. Figure~\ref{fig:hard-distributions} illustrates the transformation.

	\paragraph{Adversary matrix construction.}
	The general adversary bound allows a real matrix $\Gamma$ whose rows are positive inputs and whose columns are negative inputs, and gives
	\[
	 Q_2(\operatorname{gc}_{G_q})
	 =\Omega\!\left(\frac{\|\Gamma\|}{\max_v\|\Gamma\circ\Delta_v\|}\right).
	\]
	Here $\Gamma\circ\Delta_v$ retains an entry only when its two inputs differ at the queried vertex $v$. We first construct a matrix $B$ with the same row and column sets. For a positive input $y$ whose unique marked edge is $a\ell$, the entry $B_{y,x}$ is nonzero exactly when $x_a=1$ and planting $a\ell$ in $x$ produces $y$. Thus, each row collects the negative inputs that could have produced its positive input by planting.

	Let $\mu^-(x)$ be the probability of $x$ under the negative distribution, and let $\mu^{a\ell}(y)$ be the probability of $y$ under the positive distribution for the fixed edge $a\ell$. Write $\operatorname{plant}_{a\ell}(x)$ for the operation that marks $\ell$ and unmarks its other points. Define

	\[
	 B_{y,x}=
	 \begin{cases}
	 \displaystyle\sqrt{\frac{\mu^-(x)}{r(1-r)\,\mu^{a\ell}(y)}},
	 &x_a=1\text{ and }y=\operatorname{plant}_{a\ell}(x),\\[6pt]
	 0,&\text{otherwise}.
	 \end{cases}
	\]

	The weights account for the probabilities of the inputs. Since a point is marked with probability $r(1-r)$ and the positive distribution is obtained by conditioning on $x_a=1$ and then planting, we have
	\[
	 \sum_{\substack{x:\,x_a=1\\\operatorname{plant}_{a\ell}(x)=y}}\mu^-(x)
	 =r(1-r)\,\mu^{a\ell}(y).
	\]
	Consequently, applying $B$ to the unit vector $\bigl(\sqrt{\mu^-(x)}\bigr)_x$ gives, in the rows for each fixed edge $a\ell$, the vector $\sqrt{r(1-r)}\,\bigl(\sqrt{\mu^{a\ell}(y)}\bigr)_y$. Its squared norm is $r(1-r)$, since $\sum_y\mu^{a\ell}(y)=1$. There are $q^3$ disjoint groups of rows, so their squared norms add, giving

	\[
	 \|B\|\ge\sqrt{q^3r(1-r)}
	 =\Omega(q\sqrt{\log q}).
	\]
	Thus, combining the possible planted edges gives the desired numerator. The difficulty is to preserve this gain while bounding what remains of the matrix after restricting to inputs that differ at one queried vertex.

	\paragraph{The additive correction.}
For a fixed planted edge $a\ell$, a query outside $\ell$ and its incident points sees the same bit before and after planting. The corresponding entries of $B$ are therefore removed by the restriction for that query. A query to $\ell$ is different: planting always changes its bit from zero to one. For a fixed queried line $\ell$, the restriction retains contributions from all $q$ possible planted points on $\ell$. The matrix $B$ alone does not give the required denominator bound. We add a matrix $C$ to cancel a contribution shared by these choices, using the signed entries allowed by the general adversary bound.

To see what should be cancelled, consider the part of the calculation involving only line bits. Before planting, every line through $a$ other than $\ell$ must be unmarked. This condition is expressed by $\prod_{\substack{h\ni a\\h\ne\ell}}(1-x_h)$. Write each factor as $(1-r)-(x_h-r)$, where $x_h-r$ is the difference between the line bit and its mean. Expanding the product separates the constant term $(1-r)^{q-1}$ from terms containing one or more of these differences.

The constant term is the same for every possible planted point on $\ell$. It contains no additional line that could help identify which point was chosen. In the calculation for a query to $\ell$, the matrix $C$ cancels the contribution from this constant term. The relation $p(1-r)^{q-1}=r$ ensures that the two coefficients have the same magnitude.

After this cancellation, each remaining contribution in this restricted case contains a factor $x_h-r$ for some additional line $h\ne\ell$ through $a$. Once $\ell$ and $h$ are fixed, the planted point must be their intersection. Thus each such line can account for at most one choice of $a$. The number of possible planted points is now bounded by the number of line factors available in the term. The corresponding counting step for a point query uses the other geometric property: a queried point and a distinct planted point determine at most one possible planted line.

\paragraph{The multiplicative correction.}
The remaining difficulty is that a term can involve many points and lines, giving many choices in these counts. The full proof organizes the expressions in a basis whose elements are products associated with sets of points and lines, extending the line expansion above. We use a diagonal matrix $D$ to give smaller weights to basis elements involving more vertices. This compensates for the growth in the counting bound. A query can introduce or remove factors, so the weights decrease gradually to control the additional contribution caused by these changes.

The unit vector used for the numerator corresponds to the constant basis function, which contains no factors. We choose $C$ to map this vector to zero and $D$ to leave it unchanged. Consequently, adding $C$ and applying $D$ preserve the image used to prove the numerator bound. The additive correction removes the contribution shared by many planted edges, while the multiplicative correction limits the contributions involving many vertices.

	For the resulting adversary matrix, these estimates give
	\[
	 \max_v\frac{\|\Gamma\circ\Delta_v\|}{\|\Gamma\|}
	 =O\!\left(\frac1{q\sqrt{\log q}}\right).
	\]
	The adversary bound therefore yields $\Omega(q\sqrt{\log q})$ queries, or $\Omega(\sqrt{n\log n})$ since $n=2q^2$.

	\subsection{Organization}
	
	Section~\ref{sec:prelim} gives the notation and standard quantum tools. Section~\ref{sec:simple} gives the first $\Ot(n^{92/139})$ algorithm using one round of estimation and vertex selection. Section~\ref{sec:algorithm} refines these steps to prove Theorem~\ref{thm:main}. Section~\ref{sec:certificates} gives the extension to positive certificates of size two over general alphabets. Section~\ref{sec:lower-bound} proves the $\Omega(\sqrt{n\log n})$ lower bound. Appendix~\ref{sec:quantum-proofs} proves the quantum procedure lemmas. Appendices~\ref{sec:tests} and~\ref{sec:partition-proof} give the collision tests and the construction of the edge partition.
	
	\section{Preliminaries}\label{sec:prelim}
    \paragraph{Graph collision.}
    All graphs are simple and undirected. For an input $x$, let $M=\{v:x_v=1\}$ be the set of marked vertices. Write $N_G(v)$ for the neighbors of $v$ and $N_G(S)=\bigcup_{v\in S}N_G(v)$. An input is positive if it has a marked edge, and negative otherwise. On a negative input, $M$ is an independent set: no edge joins two vertices of $M$. Edge deletions retain the vertex labels and their input bits.

    We write $Q_2(f)$ for the quantum query complexity of $f$ with error at most $1/3$ on every input. Logarithms are base two, and $\Ot$, $\widetilde\Omega$, and $\widetilde\Theta$ suppress factors polynomial in $\log n$.

    \begin{definition}[Marked volume]\label{def:marked-volume}
        For $S\subseteq V(G)$, define
        \[
        \vol(G,S,x)=\sum_{v\in S\cap M}\deg_G(v),
        \qquad \vol(G,x)=\vol(G,V(G),x).
        \]
        This counts incidences between edges and marked vertices in $S$. An edge contributes once for each endpoint in $S\cap M$.
    \end{definition}

    \begin{lemma}[Counting marked volume]\label{lem:marked-volume}
        For every $S\subseteq V(G)$,
        \[
        \vol(G,S,x)=\sum_{u\in V(G)}|N_G(u)\cap S\cap M|.
        \]
        If the edge sets of $G_1,\ldots,G_t$ partition $E(G)$, then $\vol(G,S,x)=\sum_{i=1}^t\vol(G_i,S,x)$, retaining the same vertex labels.
    \end{lemma}
    \begin{proof}
        Both identities count each edge once for each endpoint in $S\cap M$.
    \end{proof}

    \begin{lemma}[Bipartite reduction]\label{lem:double}
		Graph collision on a known graph with $n$ vertices reduces to graph collision on a known bipartite graph with $2n$ labels, with one original query per query to the copied input.
	\end{lemma}
	\begin{proof}
		Replace each vertex $v$ by $v_0,v_1$, both marked by $x_v$, and each edge $uv$ by $u_0v_1$ and $v_0u_1$. A marked edge exists in the new graph exactly when it exists in the original graph.
	\end{proof}
	
	We use the bipartite reduction in the improved algorithm; the first algorithm applies directly to the input graph.

    \paragraph{Standard quantum procedures.}
    Grover search on a known set of $a$ labels uses $\Ot(\sqrt a)$ queries. We also use the following consequences of quantum counting, amplitude amplification, variable-time search, and quantum walks.

    \begin{lemma}[Counting and marked samples~\cite{BHMT,YLC}]\label{lem:quantum}
        Let a known set of $a$ labels contain $r$ marked vertices. With $\Ot(\sqrt a)$ queries, quantum counting returns an integer $\widehat r\in[0,a]$ that, except with the assigned error probability, is zero exactly when $r=0$ and otherwise satisfies
        \[
        r/2\le\widehat r\le2r,
        \qquad |\widehat r-r|=O(\sqrt r).
        \]
        Given $\widehat r>0$, a batch of $s$ independent uniform marked samples uses a fixed query limit $\Ot((s+1)\sqrt{a/\widehat r})$, reporting failure if insufficient samples are obtained. On a good estimate, sampling succeeds except with the assigned error probability, and the uniform state over marked labels can be prepared to inverse polynomial precision in $\Ot(\sqrt{a/\widehat r})$ queries.
    \end{lemma}
    
    \begin{lemma}[Combining tests~\cite{VT}]\label{lem:or}
		Suppose polynomially many tests have known polynomially bounded query limits $q_i$. Each accepts every positive input for its own edge list and rejects negative inputs satisfying its stated bound. If all these bounds hold on a global negative input, their OR can be tested in $\Ot(\sqrt{\sum_iq_i^2})$ queries, accepting every positive global input.
	\end{lemma}

    \begin{lemma}[Quantum walk search~\cite{MNRS}]\label{lem:walk}
		Consider a reversible ergodic Markov chain with spectral gap at least $\delta$. Suppose the set of accepting states has stationary probability at least $\eta$ whenever it is nonempty. Let $S$, $U$, and $C$ denote the coherent setup, update, and checking query costs, respectively. Quantum walk search decides whether an accepting state exists using
		\[
		\Ot\left(S
		+\frac1{\sqrt\eta}\left(\frac{U}{\sqrt\delta}
		+C\right)\right)
		\]
		queries.
	\end{lemma}

    Proofs and implementation details for Lemmas~\ref{lem:quantum} and~\ref{lem:or} appear in Appendix~\ref{sec:quantum-proofs}. Unless stated otherwise, samples are independent draws with replacement, sample sizes are rounded upward, and error probabilities are inverse polynomial in $n$. We write $\epsilon$ for the error probability assigned to a subroutine. Each sampling call uses its stated query limit; if sampling fails or a subroutine returns \Yes, the calling routine returns \Yes.

    \paragraph{Collision tests.}
    The following tests have fixed query limits, and their guarantees hold with the assigned error probability. Their constructions and proofs appear in Appendix~\ref{sec:tests}.

    \begin{lemma}[A cover of small marked volume]\label{lem:cover}
        Let $U$ be a known vertex cover of $G$. The procedure $\covertest(G,U,b)$ uses $\Ot(\sqrt{|U|}+\sqrt b)$ queries. It accepts every positive input and rejects negative inputs satisfying $\vol(G,U,x)\le b$.
    \end{lemma}

    \begin{lemma}[Bipartite collision test]\label{lem:bipartite}
        Let $G$ be bipartite with sides $A,B$ of sizes $a,b$.

        The procedure $\bipartitetest(G,A,B,k)$ accepts every positive input and rejects negative inputs satisfying $|M\cap B|\le k$, with query complexity
        \begin{equation}\label{eq:bipartite}
        \Ot\bigl(\sqrt a+\sqrt b+(a^2bk)^{1/6}\bigr).
        \end{equation}
    \end{lemma}

    \begin{lemma}[Few marked vertices in a cover]\label{lem:few}
        Let $G$ have at most $n$ vertices and let $U$ be a known vertex cover. There is a test using $\Ot\bigl(\sqrt n(1+k)^{1/6}\bigr)$ queries that accepts every positive input and rejects negative inputs satisfying $|M\cap U|\le k$.
    \end{lemma}

    The volume test applies to the whole graph with $U=V(G)$. For edges discarded by restricting to a vertex set, the vertices outside that set form a cover.

    \Needspace{9\baselineskip}
	\paragraph{Adversary bound.}
	We use the general adversary bound, which allows negative matrix entries. All matrix norms below are operator norms, and $\circ$ denotes entrywise multiplication.

	\begin{lemma}[General adversary bound~\cite{HLS07}]\label{lem:lower-adversary}
		Let $f:D\to\{0,1\}$, where $D\subseteq\{0,1\}^k$, and let $X\subseteq f^{-1}(0)$ and $Y\subseteq f^{-1}(1)$. For $v\in[k]$, let $\Delta_v(y,x)=\mathbf1[y_v\ne x_v]$. For every nonzero matrix $\Gamma$ with rows indexed by $Y$ and columns indexed by $X$, computing $f$ with error at most $1/3$ requires
		\[
		 \Omega\!\left(\frac{\|\Gamma\|}{\max_v\|\Gamma\circ\Delta_v\|}\right)
		\]
		queries.
	\end{lemma}
	\begin{proof}
		Apply the general adversary bound to the Hermitian matrix $\left(\begin{smallmatrix}0&\Gamma\\\Gamma^*&0\end{smallmatrix}\right)$, extended by zeros to the remaining inputs. Its norm is $\|\Gamma\|$, and its disagreement matrix at $v$ has norm $\|\Gamma\circ\Delta_v\|$.
	\end{proof}

	\section{An algorithm breaking the \texorpdfstring{$O(n^{2/3})$}{O(n\textasciicircum(2/3))} barrier}\label{sec:simple}
	Let $G=(V,E)$ be a known graph on $n$ vertices, with input $x\in\{0,1\}^{V}$. In this section, we give a quantum algorithm for graph collision with query complexity exponent strictly less than $2/3$.
	
	\begin{theorem}\label{thm:simple}
		There is a quantum algorithm that solves the graph collision problem on any $n$-vertex graph with high probability using $\Ot(n^{92/139})$ queries.
	\end{theorem}
	
	The algorithm uses estimates of the number of marked neighbors of each vertex. We first show how a bound on their error allows us to apply the collision tests from the preliminaries. We then construct the estimates from a random sample of input bits and describe the preprocessing needed to obtain a sufficiently small error. The algorithm first preprocesses the graph, then estimates the neighborhood counts, applies the threshold, and performs the two final collision tests. Algorithm~\ref{alg:simple} gives the complete procedure at the end of the section, followed by its query analysis.
	
	\paragraph{Part 1: Thresholding the neighborhood counts.}
	For $S\subseteq V(G)$, let $\ell_u(S)=|N_G(u)\cap S|$. Thus $\ell_u(M)$ is the number of marked neighbors of $u$. If these numbers were known exactly, we could search for a marked vertex $u$ with $\ell_u(M)>0$ using $O(\sqrt n)$ queries. Such a vertex exists exactly when there is a collision.
	
	Learning every count exactly is too expensive. Instead, we will find a set $S$ whose counts $\ell_u(S)$ approximate $\ell_u(M)$, with small total squared error. The set $S$ need not resemble $M$: only their neighborhood counts matter. Choose a threshold $\tau>0$ and let $P=\{u:\ell_u(S)\le\tau\}$. 
	
	Inside $P$, the estimated counts $\ell_u(S)$ are at most $\tau$, so their sum is at most $n\tau$. The estimation error bounds the difference from the sum of the actual marked neighbor counts, giving a bound on the marked volume of $G[P]$. On a negative input, a marked vertex has no marked neighbors. Each marked vertex outside $P$ therefore contributes more than $\tau^2$ to the total squared error. An error bound thus limits how many such vertices there can be.

	Figure~\ref{fig:thresholding} summarizes why thresholding makes both parts easy to test.

	\begin{lemma}[Thresholding]\label{lem:select} 
		Let $S\subseteq V(G)$, let $r=\sum_u(\ell_u(S)-\ell_u(M))^2$, and let $P=\{u:\ell_u(S)\le\tau\}$ for $\tau>0$. Then
		\[
		\vol(G[P],x)\le n\tau+\sqrt{nr}.
		\]
		On a negative input, $|M\setminus P|\le r/\tau^2$. If the input is negative and the maximum degree is at most $d$, the edges outside $G[P]$ have cover $V(G)\setminus P$ of marked volume at most $dr/\tau^2$.
	\end{lemma}
	\begin{proof}
		We first bound the marked volume inside $P$. Count each incidence between an edge of $G[P]$ and a marked vertex from the other endpoint of that edge. This gives
		\[
		\vol(G[P],x)=\sum_{u\in P}|N_G(u)\cap M\cap P|
		\le\sum_{u\in P}\ell_u(M).
		\]
		For every $u\in P$, the estimated count $\ell_u(S)$ is at most $\tau$. Separating these estimates from their errors, we obtain
		\[
		\sum_{u\in P}\ell_u(M)
		\le |P|\tau+\sum_{u\in P}|\ell_u(M)-\ell_u(S)|
		\le n\tau+\sqrt{nr}.
		\]
		The last inequality follows from Cauchy--Schwarz: the sum of the absolute errors over $P$ is at most $\sqrt{|P|}$ times the square root of their sum of squares, which is at most $\sqrt r$.
		
		Now suppose the input is negative. Every marked vertex $u$ has $\ell_u(M)=0$, whereas $u\notin P$ implies $\ell_u(S)>\tau$. Each vertex in $M\setminus P$ therefore contributes more than $\tau^2$ to $r$, giving $|M\setminus P|\le r/\tau^2$. Every edge outside $G[P]$ has an endpoint in $V(G)\setminus P$, so this set is a vertex cover of those edges. If the maximum degree is at most $d$, each marked vertex in this cover contributes at most $d$ to its marked volume. The volume is therefore at most $d|M\setminus P|\le dr/\tau^2$.
	\end{proof}
	
	Lemma~\ref{lem:cover} converts these volume bounds into query bounds for testing the two edge sets. The volume bounds are needed for rejection on negative inputs; a test accepts whenever its edges contain a collision, even if the supplied bound fails.

	\paragraph{Part 2: Estimating the neighborhood counts.}
	Take $q$ independent uniform vertices $w_1,\ldots,w_q$ and query their input bits. The natural estimate of the marked neighbor count at $u$ is
	\[
	 y_u=\frac nq\sum_{i=1}^q x_{w_i}\mathbf1[w_i\in N_G(u)].
	\]
	This estimate is unbiased: $\mathbb E[y_u]=\ell_u(M)$. However, the individual estimates may be noisy. The true counts have additional structure: they all arise from the same set $M$ of marked vertices. We use this structure by requiring our estimates to be the neighborhood counts of a single set $S$. We choose $S$ to minimize $\sum_u(\ell_u(S)-y_u)^2$, the total squared difference between its counts and the sampled values. Since $M$ is one of the allowed choices, the objective value at $S$ is no larger than at $M$. We can therefore compare the chosen set with $M$ without knowing $M$, and use this comparison to bound the error in the neighborhood counts.
	
	To expand the comparison, let $c_{vw}=|N_G(v)\cap N_G(w)|$ and define
	\[
	 z_v=\sum_{u\in N_G(v)}y_u
	     =\frac nq\sum_{i=1}^q c_{vw_i}x_{w_i}.
	\]
	Here, a sampled marked vertex $w_i$ contributes once for each common neighbor of $v$ and $w_i$. Since the samples are uniform, $\mathbb E[z_v]=\sum_w c_{vw}x_w$. Thus $z_v$ is an unbiased estimate of the common neighbor sum at $v$.
	
	Counting the contributions of pairs $u,v$ with $v\in S$ and $u\in N_G(v)$ in two ways gives
	\[
	\sum_u\ell_u(S)y_u
	=\sum_{v\in S}\sum_{u\in N_G(v)}y_u
	=\sum_{v\in S}z_v.
	\]
	Expanding the square in the least squares objective therefore gives
	\[
	\sum_u\bigl(\ell_u(S)-y_u\bigr)^2
	=\sum_u\ell_u(S)^2-2\sum_{v\in S}z_v+\sum_u y_u^2.
	\]
	Once the samples are fixed, the last term $\sum_u y_u^2$ is the same for every choice of $S$. Removing it does not change which sets minimize the objective. We may therefore minimize $\psi_z(S)=\sum_u\ell_u(S)^2-2\sum_{v\in S}z_v$ instead.
	
	Algorithm~\ref{alg:fit} gives the procedure for estimating the neighborhood counts from the sampled input bits.
	
	\begin{algorithm}[H]
		\caption{\estimateneighborhoodcounts}\label{alg:fit}
		\small
		\textbf{Input:} $(G,q)$.
		\begin{steps}
			\item Take $q$ fresh uniform samples $w_1,\ldots,w_q$ and query their input bits.
			\item Let $z_v=(n/q)\sum_{i=1}^q c_{vw_i}x_{w_i}$ for every vertex $v$.
			\item \textbf{return} a set $S$ minimizing $\psi_z(S)$ over all subsets of $V(G)$.
		\end{steps}
	\end{algorithm}
	
	The improved algorithm uses the same objective. The comparison requires only $\psi_z(S)\le\psi_z(M)$, which holds whenever $M$ is among the sets over which we minimize.
	
	\begin{lemma}[Comparison with the marked set]\label{lem:comparison}
		Let $S\subseteq V(G)$ satisfy $\psi_z(S)\le\psi_z(M)$. Then
		\begin{equation}\label{eq:comparison}
			\sum_u\bigl(\ell_u(S)-\ell_u(M)\bigr)^2
			\le 2\sum_v e_v\bigl(\mathbf1[v\in S]-x_v\bigr),
			\qquad e_v=z_v-\sum_w c_{vw}x_w.
		\end{equation}
	\end{lemma}
	\begin{proof}
		To relate the objective inequality to the error, we count in two ways the triples $(u,v,w)$ with $v\in S$, $w\in M$, and both $v$ and $w$ adjacent to $u$. Counting first by $u$, and then by $v,w$, gives
		\[
		\sum_u\ell_u(S)\ell_u(M)=\sum_{v\in S}\sum_w c_{vw}x_w.
		\]
		The same identity with $S=M$ expresses $\sum_u\ell_u(M)^2$ as $\sum_{v\in M}\sum_w c_{vw}x_w$. Substitute $z_v=\sum_w c_{vw}x_w+e_v$ into the difference of the two objectives. These identities then give
		\[
		\begin{aligned}
		\psi_z(S)-\psi_z(M)
		&=\sum_u\bigl(\ell_u(S)-\ell_u(M)\bigr)^2\\
		&\quad-2\sum_v e_v\bigl(\mathbf1[v\in S]-x_v\bigr).
		\end{aligned}
		\]
		The left side is nonpositive by hypothesis. Moving the error term to the other side proves~\eqref{eq:comparison}.
	\end{proof}
	
	Since $|\mathbf1[v\in S]-x_v|\le1$, the squared estimation error is at most $2\sum_v|e_v|$. Thus we can bound the error of the estimated neighborhood counts by bounding the errors in the sampled common neighbor sums. All common neighbor counts are known from the graph, and the minimization uses no further input queries.
	
	\paragraph{The accuracy needed from preprocessing.}
	The variance of $z_v$ depends on $\sum_{w\in M}c_{vw}^2$. For fixed $v,w$, the quantity $c_{vw}^2$ counts pairs of common neighbors, or equivalently closed walks $w,a,v,b,w$ of length four. Summing over $v$ and marked $w$ counts just the closed walks starting at marked vertices. We will use sampling and edge deletions to ensure that this count is at most $2h\,\vol(G,x)$, for a parameter $h\ge1$, and that $\vol(G,x)\le n^2/q$.
	
	\begin{lemma}[Accuracy of the neighborhood estimates]\label{lem:fit}
		For $h\ge1$, suppose $\vol(G,x)\le n^2/q$ and
		\begin{equation}\label{eq:walkcount}
			\sum_{w\in M}\sum_v c_{vw}^2\le2h\,\vol(G,x).
		\end{equation}
		Then Algorithm~\ref{alg:fit} has expected squared estimation error $O(n^2\sqrt h/q)$.
	\end{lemma}
	\begin{proof}
		Fix a vertex $v$. For a uniform sample $w$, the mean of $nc_{vw}x_w$ is $\sum_{w\in M}c_{vw}$, and its second moment is $n\sum_{w\in M}c_{vw}^2$. Thus $z_v$ is unbiased. Since it averages $q$ independent samples, its variance is
		\[
		\mathbb E[e_v^2]=\operatorname{Var}(z_v)
		=\frac1q\operatorname{Var}(nc_{vw}x_w)
		\le\frac nq\sum_{w\in M}c_{vw}^2.
		\]
		Summing over $v$ and using the two hypotheses gives
		\[
		\mathbb E\sum_v e_v^2
		\le\frac nq\sum_{w\in M}\sum_v c_{vw}^2
		\le\frac{2nh}q\,\vol(G,x)
		\le\frac{2n^3h}{q^2}.
		\]
		
		By Lemma~\ref{lem:comparison}, the squared error in the neighborhood counts is at most $2\sum_v|e_v|$. Applying Cauchy--Schwarz first to each expectation and then to the sum over vertices bounds its expectation by
		\[
		2\sum_v\mathbb E|e_v|
		\le2\sum_v\sqrt{\mathbb E[e_v^2]}
		\le2\sqrt{n\,\mathbb E\sum_v e_v^2}
		=O(n^2\sqrt h/q).
		\]
		If $G$ was obtained by earlier preprocessing, condition on any history for which the two hypotheses hold. The graph is then fixed, and the fresh samples have the distribution used above, so the same bound holds conditionally.
	\end{proof}
	
	\paragraph{Part 3: Reducing marked volume and maximum degree.}
	We have analyzed estimation on a graph satisfying the volume and closed walk bounds in Lemma~\ref{lem:fit}. We now describe the preprocessing that establishes these properties. In this part, we first reduce the marked volume by sampling, then test and delete edges incident to vertices of degree greater than $d$. The degree bound turns the bound on the number of marked vertices outside $P$ into the volume bound in Lemma~\ref{lem:select}, and keeps neighborhoods small for the graphs tested in Part 4. Both preprocessing procedures precede the fresh samples used to estimate the neighborhood counts.
	
	We begin by reducing the marked volume. Sample vertices, query their input bits, and search the union of the neighborhoods of the marked samples. A marked vertex in this union gives a collision with a sampled marked vertex. If the search finds none, delete all edges incident to this union.
	
	The reason for this deletion is simple: a vertex with many marked neighbors is likely to have one of them sampled. Such a vertex then belongs to the searched union and is isolated. Thus the remaining vertices have few marked neighbors, and their total count bounds the marked volume. Algorithm~\ref{alg:sample} gives the procedure.
	
	\begin{algorithm}[H]
		\caption{\sampleneighborhoods}\label{alg:sample}
		\small
		\textbf{Input:} $(G,q)$, with $q\ge1$. 
		\begin{steps}
			\item Take $\Theta(q\log(n/\epsilon))$ uniform vertex samples and query their input bits.
			\item Let $Z$ be the marked sampled labels.
			\item Search $N_G(Z)$ for a marked vertex; return \Yes\ if one is found.
			\item Delete every edge incident to $N_G(Z)$ and return $G$.
		\end{steps}
	\end{algorithm}
	
	\begin{lemma}[Neighborhood sampling]\label{lem:sample}
		For $q\ge1$, Algorithm~\ref{alg:sample} has cost $\Ot(q+\sqrt n)$ queries. Except with probability $\epsilon$, every remaining vertex has at most $n/q$ marked neighbors and the remaining marked volume is at most $n^2/q$. If a collision exists, the algorithm either detects one or leaves one in the remaining graph, unless the Grover search fails.
	\end{lemma}
	\begin{proof}
		Let $t=\Theta(q\log(n/\epsilon))$ be the number of samples. Querying their input bits costs $t$ queries. Once these bits are known, the set $N_G(Z)$ is known, so Grover search over it costs $\Ot(\sqrt n)$ queries.
		
		Fix a vertex $u$ with more than $n/q$ marked neighbors before deletion. It lies outside $N_G(Z)$ exactly when none of its marked neighbors is sampled. By independence, this event has probability
		\[
		\left(1-\frac{\ell_u(M)}n\right)^t
		\le\exp\left(-\frac{t\ell_u(M)}n\right)
		\le\exp(-t/q)
		\le\epsilon/n,
		\]
		for a sufficiently large constant in the choice of $t$. A union bound over the vertices shows that, except with probability $\epsilon$, every vertex outside $N_G(Z)$ has at most $n/q$ marked neighbors. Deleting edges cannot increase this count, and every vertex inside $N_G(Z)$ becomes isolated. In the resulting graph, counting in two ways therefore gives $\vol(G,x)=\sum_u|N_G(u)\cap M|\le n^2/q$.
		
		Finally, if a deleted edge is a collision, one of its marked endpoints lies in $N_G(Z)$. If the Grover search succeeds, it finds a marked vertex in this set, which forms a collision with a sampled marked neighbor in $Z$. If no deleted edge is a collision, any existing collision remains in the graph.
	\end{proof}
	
	\paragraph{Bounding the maximum degree.}
	On the event guaranteed by Lemma~\ref{lem:sample}, the remaining graph has marked volume at most $n^2/q$. Let $H=\{u:\deg_G(u)>d\}$. Each marked vertex in $H$ contributes more than $d$ to the volume, so $H$ contains at most $n^2/(qd)$ marked vertices. Algorithm~\ref{alg:simple} tests the edges incident to $H$ using Lemma~\ref{lem:few}, then deletes them if the test rejects. The maximum degree is then at most $d$, and the volume bound is preserved.
	
	\paragraph{Part 4: Bounding closed walks of length four.}
	The volume bound alone does not control the squared common neighbor counts in~\eqref{eq:walkcount}. This last preprocessing step supplies that bound, which controls the variance of the estimates $z_v$. Fix the first directed edge $wa$ of a closed walk $w,a,v,b,w$. The edge $vb$ joins $N_G(a)$ to $N_G(w)$, and choosing its orientation determines the whole walk.
	
	Whenever an edge $wa$ in the current graph has at least $h$ edges between $N_G(a)$ and $N_G(w)$, record and delete those edges. Repeat until no such edge remains. Each remaining directed edge then starts at most $2h$ closed walks of length four: there are fewer than $h$ choices for the middle edge and at most two orientations for each. There are $\vol(G,x)$ directed edges starting at marked vertices, so the total number of these walks is at most $2h\,\vol(G,x)$. This is exactly the bound needed in~\eqref{eq:walkcount}.

	We use Grover search over the recorded graphs to find one containing a collision, using Lemma~\ref{lem:few} to check a graph. Each recorded graph has at most $2d$ vertices and can be checked by a quantum walk. The walks of length four above are ordinary graph walks used for counting; the quantum walk is the procedure used for collision testing. A larger threshold $h$ gives fewer recorded graphs, but allows a larger estimation error. Algorithm~\ref{alg:basic} gives the procedure.
	
	Figure~\ref{fig:extraction} illustrates how extraction separates collision testing from estimation on the remaining graph.

	\begin{algorithm}[H]
		\caption{\extractneighborhoodedges}\label{alg:basic}
		\small
		\textbf{Input:} A graph $G$ of maximum degree at most $d$, and a threshold $h\ge1$.
		\begin{steps}
			\item Initialize an empty list of recorded graphs.
			\item \textbf{while} an edge $uv$ has at least $h$ edges between $N_G(u)$ and $N_G(v)$:
			\item \ind Record those edges as a graph on their endpoints.
			\item \ind Delete the recorded edges from $G$.
			\item \textbf{end while}
			\item Search the recorded graphs using Lemma~\ref{lem:few}, with count bound $2d$.
			\item Return \Yes\ if a test accepts; otherwise return $G$.
		\end{steps}
	\end{algorithm}
	
	\begin{lemma}[Edge extraction and closed walks]\label{lem:basic}
		Algorithm~\ref{alg:basic} costs $\Ot(\sqrt{n/h}\,d^{7/6})$ queries. Its output satisfies~\eqref{eq:walkcount}.
		If any deleted edge is a collision, Algorithm~\ref{alg:basic} returns \Yes\ with high probability.
	\end{lemma}
	\begin{proof}
		We first prove the walk bound in the remaining graph. For fixed $v,w$, each ordered pair of common neighbors $a,b\in N_G(v)\cap N_G(w)$ gives the closed walk $w,a,v,b,w$. Conversely, such a walk determines $v,a,b$ once its initial vertex $w$ is fixed. Thus $\sum_{w\in M}\sum_v c_{vw}^2$ counts exactly the closed walks of length four starting at marked vertices.
		
		Count these walks instead by their first directed edge $wa$, with $w$ marked. The edge $vb$ joins $N_G(a)$ to $N_G(w)$. At termination there are fewer than $h$ such edges. Each undirected edge gives at most two choices for the ordered pair $(v,b)$, and each choice determines the whole walk. There are therefore at most $2h$ walks for each first directed edge. The number of first directed edges with a marked initial vertex is $\sum_{w\in M}\deg_G(w)=\vol(G,x)$, proving~\eqref{eq:walkcount}. This argument also counts walks with repeated vertices: neither the ordered pair of common neighbors nor the description by a first edge requires distinct vertices.
		
		For the query bound, each recorded graph is formed from edges between two neighborhoods of size at most $d$, so it has at most $2d$ vertices. The recorded edge sets are disjoint, each has at least $h$ edges, and the original graph has at most $nd$ edges. Hence at most $nd/h$ graphs are recorded. Using all vertices of a recorded graph as its cover, Lemma~\ref{lem:few} with count bound $2d$ tests it in $\Ot(\sqrt d(1+2d)^{1/6})=\Ot(d^{2/3})$ queries. Grover search over the recorded graphs then costs $\Ot(\sqrt{nd/h}\,d^{2/3})=\Ot(\sqrt{n/h}\,d^{7/6})$ queries. Every deleted edge belongs to a recorded graph. If such an edge is a collision, the search over the recorded graphs returns \Yes\ with high probability.
	\end{proof}
	
	\paragraph{The complete algorithm.}
	The preprocessing steps supply exactly the two hypotheses needed to estimate the neighborhood counts: small marked volume and few closed walks of length four from marked vertices. A fresh sample then gives approximate neighborhood counts, and thresholding divides the remaining edges into two parts handled by the same volume test. Algorithm~\ref{alg:simple} combines these steps.
	
	Choose parameters $q,d,h\ge1$ and $\tau>0$, with $q$ an integer. Let $r$ be a sufficiently large constant multiple of $n^2\sqrt h/q$, and let $b=n\tau+\sqrt{nr}+dr/\tau^2$. Lemma~\ref{lem:fit} and Markov's inequality make $r$ a valid error bound with high constant probability. Lemma~\ref{lem:select} then makes $b$ a valid marked volume bound for both final tests.
	
	Return \Yes\ as soon as any search or collision test returns \Yes.
	
	\begin{algorithm}[H]
		\caption{First algorithm}\label{alg:simple}
		\small
		\textbf{Input:} $(G,q,d,h,\tau)$, with $b$ defined above.
		\begin{steps}
			\item $G\gets\sampleneighborhoods(G,q)$.
			\item Let $H=\{u:\deg_G(u)>d\}$.
			\item Test its incident edges by Lemma~\ref{lem:few}, with count bound $n^2/(qd)$.
			\item Delete the edges incident to $H$.
			\item $G\gets\extractneighborhoodedges(G,d,h)$.
			\item $S\gets\estimateneighborhoodcounts(G,q)$.
			\item Let $P=\{u:\ell_u(S)\le\tau\}$ and $F=(V(G),E(G)\setminus E(G[P]))$.
			\item Call $\covertest(F,V(G)\setminus P,b)$.
			\item \textbf{return} $\covertest(G[P],P,b)$.
		\end{steps}
	\end{algorithm}
	
	\begin{proof}[Proof of Theorem~\ref{thm:simple}]
		Suppose first that the input is negative. Condition on neighborhood sampling giving marked volume at most $n^2/q$. The set $H$ then contains at most $n^2/(qd)$ marked vertices, so the high degree test has a valid count bound and rejects except with its assigned error probability. Deleting its incident edges leaves maximum degree at most $d$, and further edge deletions preserve the volume bound. Lemma~\ref{lem:basic} establishes~\eqref{eq:walkcount}. The fresh samples therefore satisfy the hypotheses of Lemma~\ref{lem:fit}. By Markov's inequality and the choice of $r$, their squared estimation error is at most $r$ with conditional probability at least $99/100$.
		
		On this event, Lemma~\ref{lem:select} bounds the marked volume of $G[P]$ by $n\tau+\sqrt{nr}$ and that of the cover $V(G)\setminus P$ for the other edges by $dr/\tau^2$. Both quantities are at most $b$. Thus both final tests have valid volume bounds and reject except with their assigned error probabilities. The other tests also reject on this negative input when their stated guarantees hold.
		
		Now suppose the input is positive and fix a collision edge. If it is deleted during preprocessing, either an endpoint was searched in the neighborhood sampling step, or the edge was included in the high degree test or a recorded graph. The corresponding search or collision test accepts except with its assigned error probability. If the edge survives preprocessing, it either lies inside $G[P]$ or has an endpoint outside $P$, so one of the final tests accepts. The acceptance guarantees of these tests hold on positive inputs even when the supplied count or volume bound fails. Consequently, this argument does not require the estimation error to be small on a positive input. Taking the assigned errors sufficiently small and applying a union bound gives total error at most $1/3$.
		
		It remains to choose the parameters and bound the cost. Combining the sampling, high degree, extraction, and final test costs gives
		\[
		\Ot\left(
		q+\sqrt n\left(1+\frac{n^2}{qd}\right)^{1/6}
		+\sqrt{\frac nh}\,d^{7/6}
		+\sqrt{n\tau}
		+\frac{n^{3/4}h^{1/8}}{q^{1/4}}
		+\frac{n\sqrt d\,h^{1/4}}{\sqrt q\,\tau}
		\right)
		\]
		queries. The last three terms bound $\sqrt b$ after substituting $r=O(n^2\sqrt h/q)$.
		
		Increasing $q$ uses more sample queries but reduces both the marked volume after neighborhood sampling and the estimation error. A smaller $d$ makes the extracted graphs cheaper to test and improves the volume bound outside $P$, but leaves more vertices for the high degree test. A larger $h$ gives fewer extracted graphs, at the cost of a weaker estimation guarantee. Finally, increasing $\tau$ improves the volume bound outside $P$ while weakening the bound inside $P$. We balance these costs by choosing
		$q=\lceil n^{92/139}\rceil$, $d=\lceil n^{51/139}\rceil$, $h=n^{74/139}$, and $\tau=n^{45/139}$.
		Then $r=O(n^{223/139})$ and $b=O(n^{184/139})$.
		Every term in the displayed query bound is $O(n^{92/139})$.
		The total is $\Ot(n^{92/139})$, with the displayed query limits on every execution. Repeating the algorithm $O(\log n)$ times and taking the majority gives the stated high probability guarantee.
	\end{proof}
	
	Minimization over vertex sets may require exponential classical computation. The comparison proofs also hold for weights in $[0,1]$, giving convex quadratic minimizations with the same query bounds.
	
	\Needspace{14\baselineskip}
	\section{An improved algorithm}\label{sec:algorithm}

	In this section, we show the full quantum algorithm for the graph collision problem, proving \Cref{thm:main}. We pick a degree threshold $1 \leq d \leq n$, and we divide the graph collision problem into two parts. In~\Cref{subsec:high-degree}, we look for a collision between two vertices where at least one has degree greater than $d$. Then, in~\Cref{subsec:low-degree}, we look for a collision between two low-degree vertices, i.e., for which both degrees are at most $d$. We finally balance the choice of $d$ in \Cref{subsec:full-algo}.

    Before turning to either of these algorithms, we introduce an algorithmic component that is important for both approaches.

    \paragraph{Marked neighborhood sampling.} We introduce a useful algorithmic building block. In a graph $G$, suppose a set $S \subseteq V(G)$ of size $a$ contains $r>0$ marked vertices. We can sample a marked vertex uniformly at random using $\Ot(\sqrt{a/r})$ queries. Taking $\Theta(s\log(n/\epsilon))$ independent samples, with replacement, therefore costs $\Ot(s\sqrt{a/r})$ queries for $s\ge1$. Let $Z$ be the set of sampled vertices. We then search its neighborhood for a marked vertex using $\Ot(\sqrt n)$ queries.

	If the search finds no collision, delete every edge incident to $N_G(Z)$. Any vertex with more than $r/s$ marked neighbors in $S$ belongs to $N_G(Z)$ with high probability and is therefore isolated by this deletion. A union bound shows that every remaining vertex has at most $r/s$ marked neighbors in $S$. Every deleted collision has a marked endpoint in the searched neighborhood, so it would have been detected, except with the search error.

	Algorithm~\ref{alg:marked-sample} estimates $r$ and then performs this sampling and deletion.
	
	\begin{algorithm}[H]
		\caption{\samplemarked}\label{alg:marked-sample}
		\small
		\textbf{Input:} A graph $G$, source $S \subseteq V(G)$, and threshold $\theta>0$.
		\begin{steps}
			\item Estimate $r=|M\cap S|$ by $\widehat r$ using Lemma~\ref{lem:quantum}.
			\item If $\widehat r=0$, return $G$.
			\item Obtain $\Theta((1+\widehat r/\theta)\log(n/\epsilon))$ uniform marked vertices from $S$.
			\item Let $Z$ be the sampled vertices.
			\item Search $N_G(Z)$ for a marked vertex; return \Yes\ if one is found.
			\item Delete every edge incident to $N_G(Z)$ and return $G$.
		\end{steps}
	\end{algorithm}
	
	\begin{lemma}[Marked neighbor bound]\label{lem:marked-neighbors}
		Let $S$ be a fixed source set of size $a$, let $r=|M\cap S|$, and let $\theta>0$. Conditional on a good count estimate, Algorithm~\ref{alg:marked-sample} uses $\Ot(\sqrt n+\sqrt{ar}/\theta)$ queries. Except with the assigned error, every remaining vertex has at most $\theta$ marked neighbors in $S$. A collision is detected or survives, except with the search error.
	\end{lemma}
	\begin{proof}
		For a good nonzero estimate, the true count lies between $\widehat r/2$ and $2\widehat r$. A vertex with more than $\theta$ marked source neighbors is missed by all samples with probability at most $\epsilon/n$. A union bound proves the neighbor bound. Any deleted collision has a marked endpoint in $N_G(Z)$ and is detected by the preceding search.
		
		Counting costs $\Ot(\sqrt a)$, sampling costs $\Ot(\sqrt{a/\widehat r}+\sqrt{a\widehat r}/\theta)$, and the search costs $\Ot(\sqrt n)$. On a good nonzero estimate, $1\le\widehat r\le2r$, proving the stated query bound. The sampling limit is fixed once the estimate is obtained. On a good zero estimate the source has no marked vertices, so no operation is needed.
	\end{proof}

	\subsection{Case I: High-degree vertices}
    \label{subsec:high-degree}

	The core idea for high-degree vertices is as follows. Let $H=\{v:\deg_G(v)>d\}$, and let $r$ be its number of marked vertices. If $r\le(n/d)^{3/2}$, we can cheaply test for a collision incident to $H$. Otherwise, we can cheaply take $\Theta((n/d)\log(n/\epsilon))$ uniform marked samples from $H$ and search their neighborhood. Algorithm~\ref{alg:marked-sample} performs this sampling and search, then deletes the edges incident to the searched neighborhood.
	
	With high probability, every vertex in the remaining graph has at most $rd/(2n)$ marked neighbors in the old set $H$. Counting in two ways bounds the sum of the remaining degrees of marked vertices in $H$ by $rd/2$. Each vertex whose degree still exceeds $d$ contributes more than $d$ to this sum, so at most $r/2$ such marked vertices remain. After logarithmically many iterations, their number is at most $(n/d)^{3/2}$. Algorithm~\ref{alg:cap} gives this degree reduction procedure.

	Maintain an upper bound $R$ on the number of marked vertices in the current high-degree set $H=\{u:\deg_G(u)>d\}$, starting with $R=n$. While $R>4k$, apply Algorithm~\ref{alg:marked-sample} with threshold $Rd/(8n)$. Counting in two ways then bounds the number of remaining marked high-degree vertices by $R/8$, so we may halve $R$ and repeat. Once $R\le4k$, test and delete the edges incident to $H$. Only the sampling routine estimates marked counts internally.

	\begin{algorithm}[H]
		\caption{\capdegrees}\label{alg:cap}
		\small
		\textbf{Input:} A graph $G$ and target degree bound $d$, with $1\le d\le n$.
		Abort the routine and output \No\ if its total query count exceeds a sufficiently large budget $\Ot(n^{3/4}/d^{1/4})$.
		\begin{steps}
			\item Let $k=\lceil(n/d)^{3/2}\rceil$ and $R=n$.
			\item \textbf{while} $R>4k$:
			\item \ind Let $H=\{u:\deg_G(u)>d\}$.
			\item \ind $G\gets\samplemarked(G,H,Rd/(8n))$.
			\item \ind Let $R\gets R/2$.
			\item \textbf{end while}
			\item Let $H=\{u:\deg_G(u)>d\}$.
			\item Test the edges incident to $H$ by Lemma~\ref{lem:few}, with count bound $4k$.
			\item Delete those edges and return $G$.
		\end{steps}
	\end{algorithm}
	
	\begin{lemma}[Degree reduction]\label{lem:cap}
		For $1\le d\le n$, Algorithm~\ref{alg:cap} has query cost
		\[
		\Ot\left(\frac{n^{3/4}}{d^{1/4}}\right).
		\]
		On a good negative execution it returns a graph of maximum degree at most $d$, hence marked volume at most $nd$. On a positive input it detects a collision or preserves one, except with the assigned error.
	\end{lemma}
	\begin{proof}
		Initially $|M\cap H|\le n=R$. Suppose this invariant holds before a sampling call, and condition on its success guarantees. Let $G'$ be the returned graph and $H'=\{u:\deg_{G'}(u)>d\}$. Since $H'\subseteq H$, Lemma~\ref{lem:marked-neighbors} and counting in two ways give
		\[
		d|M\cap H'|\le\sum_u|N_{G'}(u)\cap M\cap H|
		\le n\frac{Rd}{8n}=\frac{Rd}{8}.
		\]
		Thus halving $R$ preserves the invariant. After $O(\log n)$ iterations, $R\le4k$, making the final cover test valid on a good negative execution. Deleting its tested edges leaves maximum degree at most $d$. Every removed collision is covered by a neighborhood search or the final test.

		Writing $r=|M\cap H|\le R$ and $|H|\le n$, each sampling call costs, on a good count estimate,
		\[
		\Ot\left(\sqrt n+\frac{\sqrt{nr}}{Rd/(8n)}\right)
		\le\Ot\left(\sqrt n+\frac{n^{3/2}}{d\sqrt R}\right).
		\]
		Since $R$ halves, the second terms sum geometrically to $\Ot(n^{3/2}/(d\sqrt k))$. The final test costs $\Ot(\sqrt n(1+k)^{1/6})$, giving total cost
		\[
		\Ot\left(\sqrt n(1+k)^{1/6}+\frac{n^{3/2}}{d\sqrt k}\right)
		=\Ot(n^{3/4}/d^{1/4}).
		\]
		If the loop is skipped, only the final test is needed and the same bound holds. This runtime bound holds on every input whenever the internal counting and sampling guarantees hold, or a collision is detected earlier. A union bound over the $O(\log n)$ calls makes their failure probability inverse polynomially small with any desired fixed exponent. Choose the global budget sufficiently large to cover every such good execution. The global cutoff then enforces the stated worst-case query bound, and returning \No\ on timeout adds only this negligible probability to the error on positive inputs. It introduces no additional acceptance on negative inputs.
	\end{proof}

	\subsection{Case II: Low-degree vertices}
    \label{subsec:low-degree}

	We now turn to the more complicated setting where we look for a collision between two vertices with small degree, i.e., $\deg(v) \leq d$. In other words, we simply remove all high-degree vertices from $G$, and look for a collision in the remaining graph.

    The degree bound gives $\vol(G,x)\le nd$. Our goal is to test and delete edges until the marked volume is small enough for the cover test of Lemma~\ref{lem:cover}. If $\ell$ is its current upper bound, that test uses $\Ot(\sqrt n+\sqrt\ell)$ queries. We therefore start with $\ell=nd$ and construct a routine that halves $\ell$ on each iteration. Algorithm~\ref{alg:bounded-degree} gives this procedure.

    We first analyze the minimization and thresholding steps, then give the graph preprocessing and estimation routines that supply their hypotheses. We combine these routines at the end of the section.

	\paragraph{Part 1: Estimating the neighborhood counts.}

    First consider a bipartite graph $G$ with sides $A$ and $B$. For vertices $u,v$ on the same side, let $c_{uv}=|N_G(u)\cap N_G(v)|$. Let $C_A$ and $C_B$ be the corresponding matrices indexed by $A$ and $B$, respectively, and let $C=C_A\oplus C_B$. Suppose that $z$ approximates $C\mathbf{1}_M$ with $\|C\mathbf{1}_M-z\|_\infty\le\eta$. The core idea is to approximately learn the set of vertices that have few connections to marked vertices, i.e., to find an approximation $P$ of all vertices $v \in V(G)$ that have at most $\tau$ marked neighbors. Then, the marked volume of $G[P]$ is small, because we exactly remove those vertices that are adjacent to many marked vertices. On the other hand, for a negative input, the marked volume of $G$ with $V(G) \setminus P$ as a source set is small as well, since all the neighbors of marked vertices are themselves unmarked. We make this intuition precise in the following lemma.
	
	\Needspace{10\baselineskip}
	\begin{lemma}[Separate thresholds]\label{lem:two-sides}
		Let $G$ be a bipartite graph with $m$ edges, with vertex sets $A$ and $B$ on both sides of the bipartition, sizes $a$ and $b$, degree upper bounds $d_A$ and $d_B$, and such that $|M \cap A| \leq ap$ and $|M \cap B| \leq bp$. Suppose that $\|z - C\mathbf{1}_M\|_{\infty} \leq \eta$. Let $S$ minimize $\psi_z(S)$ over subsets satisfying $|S\cap A|\le ap$ and $|S\cap B|\le bp$. Define $P=\{u\in A:\ell_u(S)\le\tau_A\}\cup\{u\in B:\ell_u(S)\le\tau_B\}$.
		Then
		\[
		\vol(G[P],x)\le a\tau_A+b\tau_B+4\sqrt{abp\eta}.
		\]
		On a negative input, the outside cover has marked volume at most
		$4p\eta(d_A b/\tau_A^2+d_B a/\tau_B^2)$.
	\end{lemma}
    
	\begin{proof}
		Replacing $S\cap B$ by $M\cap B$ preserves feasibility. Minimality and the comparison~\eqref{eq:comparison} therefore bound the squared estimation error on $A$ by
		\[
		2\sum_{v\in B}e_v(\mathbf{1}[v\in S]-x_v)
		\le2\eta|(S\triangle M)\cap B|
		\le2\eta\bigl(|S\cap B|+|M\cap B|\bigr)
		\le4bp\eta,
		\]
		where $e_v=z_v-\sum_w c_{vw}x_w$. Exchanging the sides gives error at most $4ap\eta$ on $B$.

		Apply the thresholding argument of Lemma~\ref{lem:select} separately on each side. Cauchy--Schwarz bounds the retained marked volume by
		\[
		a\tau_A+b\tau_B+\sqrt{a\cdot4bp\eta}+\sqrt{b\cdot4ap\eta}
		=a\tau_A+b\tau_B+4\sqrt{abp\eta}.
		\]
		On a negative input, every marked vertex has no marked neighbors. A marked vertex outside $P$ thus has estimation error greater than its side's threshold. There are at most $4bp\eta/\tau_A^2$ such vertices on $A$ and $4ap\eta/\tau_B^2$ on $B$. Multiplying by the respective degree bounds proves the outside-cover bound.
	\end{proof}

	The two volume bounds specify the accuracy needed from estimation. The retained volume must decrease, while the outside cover must remain cheap to test. We choose the thresholds and estimation accuracy together in the final parameter analysis. We now construct estimates satisfying these requirements.

	\paragraph{Part 2: Graph partitioning.} To simplify the problem, we find an edge partition of the graph into polylogarithmically many bipartite subgraphs with roughly uniform degrees on both sides of the bipartition. If for one of the subgraphs we label the sides $A$ and $B$, then we have degree bounds $d_A \leq d$ and $d_B \leq d$ such that for $v \in A$ and $w \in B$,
	\[\frac{d_A}{\gamma} \leq \deg(v) \leq d_A, \qquad \text{and} \qquad \frac{d_B}{\gamma} \leq \deg(w) \leq d_B,\]
	for a multiplicative factor $\gamma$ that is polylogarithmic in $n$. This construction does not require any queries to the marked set, and hence does not incur any overhead in the resulting query complexity. Lemma~\ref{lem:partition} states the partition guarantee, and Algorithm~\ref{alg:partition} in Appendix~\ref{sec:partition-proof} gives the construction. 

    Since the marked volume is additive over edge-disjoint graphs, we observe that for any bipartite subgraph, we have the same upper bound $\ell$ on the marked volume. Since we simultaneously have a lower bound on the degrees, this translates into an upper bound on the number of marked vertices on either side of the bipartition. Indeed, we find that
	\[|M \cap A| \leq \frac{\ell\gamma}{d_A} \in \widetilde{O}\left(\frac{\ell}{d_A}\right), \qquad \text{and} \qquad |M \cap B| \leq \frac{\ell\gamma}{d_B} \in \widetilde{O}\left(\frac{\ell}{d_B}\right).\]
    Indeed, on later iterations where $\ell$ is smaller, this sparsity bound becomes more stringent, making the routines that exploit this sparsity more efficient.

    Next, we run marked neighbor sampling on both sides of the bipartition individually, to obtain upper bounds $\theta_A$ and $\theta_B$ with high probability.

	\begin{lemma}
		Let $G$ be bipartite with sides $A,B$, whose degrees lie in $[d_A/\gamma,d_A]$ and $[d_B/\gamma,d_B]$, respectively. Suppose $\vol(G,x)\le\ell$, and let $\theta_A,\theta_B>0$. Applying Algorithm~\ref{alg:marked-sample} with source sets $A$ and $B$ and thresholds $\theta_A$ and $\theta_B$, respectively, uses, conditional on good count estimates,
		\[\widetilde{O}\left(\sqrt{|A|} + \sqrt{|B|} + \frac{\sqrt{|A| \cdot \frac{\gamma\ell}{d_A}}}{\theta_A} + \frac{\sqrt{|B| \cdot \frac{\gamma\ell}{d_B}}}{\theta_B}\right)\]
		queries. Except with the assigned error probability, the routine detects a collision or returns a graph in which every vertex in $A$ has at most $\theta_B$ marked neighbors and every vertex in $B$ has at most $\theta_A$ marked neighbors.
	\end{lemma}

	\begin{proof}
		The degree lower bounds give $|M\cap A|\le\gamma\ell/d_A$ and $|M\cap B|\le\gamma\ell/d_B$. Apply Lemma~\ref{lem:marked-neighbors} on both sides and substitute these count bounds into its cost. Later edge deletions preserve the marked neighbor bounds.
	\end{proof}

	\paragraph{Part 3: $C$-preconditioning.} We now study the matrix $C_A$, indexed by vertices from $A$, where $c_{uv} = |N_G(u) \cap N_G(v)|$. Thus $C_A$ is the product of the bipartite adjacency matrix with its transpose. That is, every entry $c_{uv}$ contains the number of walks of length two between $u$ and $v$. Since the degrees of $u$ and $v$ are bounded by $d_A$, we obtain that $c_{uv}$ is trivially bounded by $d_A$ as well.

	Let's consider a single row of $C_A$, indexed by the vertex $u \in A$. Intuitively, if we consider a subset $R \subseteq A$, such that for all $v \in R$ the entries $c_{uv}$ are large, then there must be many edges between $R \subseteq A$ and $N_G(u) \subseteq B$. In fact, the number of edges between $N_G(u)$ and $R$ is
	\[c_{u,R} := \sum_{v \in R} c_{uv}.\]
	We refer to $R$ as a \textit{dense} region if $c_{u,R}^2/|R| \geq s_A\theta_A d_A/(2\log(n))$. The idea is to search for such dense regions first, and check for collisions in these regions separately.

	Algorithm~\ref{alg:extract} makes this intuition precise.
	
	\begin{algorithm}[H]
		\caption{\prepareestimates}\label{alg:extract}
		\small
		\textbf{Input:} A bipartite graph $G$, with vertex sets $A$ and $B$, such that for all $v \in A$ and $w \in B$, $\deg(v) \leq d_A$, $\deg(w) \leq d_B$, and the number of marked neighbors is at most $\theta_B$ and $\theta_A$, respectively. Let $s_A\ge1$ and $\delta>0$.
		\begin{steps}
			\item Initialize an empty list of recorded graphs.
			\item \textbf{for each} row $u\in A$, in any fixed order:
			\item \ind Let $A' := \{v \in A : c_{uv} \geq s_A\}$, and sort it such that $c_{uv}$ is decreasing.
			\item \ind Find the longest initial segment $R$ such that $c_{u,R}^2/|R| \geq s_A\theta_A d_A/(2\log(n))$. If none exist, continue.
			\item \ind If $c_{u,R} \leq \delta$, continue.
			\item \ind Record the edges between $R$ and $N_G(u)$, then delete them.
			\item \textbf{end for}
			\item Test the recorded graphs using \bipartitetest{} and Lemma~\ref{lem:or}.
			\item \textbf{return} $G$.
		\end{steps}
	\end{algorithm}
	
	\begin{lemma}[Extraction and coefficient bounds]\label{lem:extract}
		Let $G$ satisfy the input conditions of Algorithm~\ref{alg:extract}, and let $m=|E(G)|$. For any $s_A\ge1$ and $\delta>0$, the algorithm records at most one graph per row $u\in A$ and uses
		\begin{equation}\label{eq:extractcost}
			\Ot\left(\sqrt m\left[
			s_A^{-1/3}(\theta_B/\theta_A)^{1/6}
			+s_A^{-1/2}+\sqrt{d_A/\delta}\right]\right)
		\end{equation}
		queries. On a negative input, every row $u\in A$ of the output graph has a set $E_u\subseteq A$ such that
		\begin{equation}\label{eq:coefficients}
		 \sum_{v\in M\cap E_u}c_{uv}\le\delta,
		 \qquad
		 \sum_{v\in M\cap(A\setminus E_u)}c_{uv}^2\le 2s_A\theta_A d_A=:v_A.
		\end{equation}
		Here $c_{uv}$ denotes the common neighbor count in the output graph. On a positive input, a collision is detected or survives, except with the assigned error. Running the routine with $A$ and $B$ exchanged gives the corresponding bounds for rows in $B$, with $v_B=2s_B\theta_B d_B$; subsequent edge deletions preserve these coefficient bounds.
	\end{lemma}

	\begin{proof}
		Put $T=s_A\theta_A d_A/(2\log n)$. Consider the moment when row $u$ is processed. If its longest prefix satisfying $c_{u,R}^2/|R|\ge T$ has $c_{u,R}\le\delta$, then every prefix of sum greater than $\delta$ fails this inequality. The same holds if no such prefix exists.

		Otherwise the algorithm extracts $R$. If $u\in R$, every edge incident to $u$ is deleted, so its row becomes zero and remains zero. If $u\notin R$, bipartiteness implies that extraction leaves $N_G(u)$ unchanged, zeros the coefficients indexed by $R$, and leaves the other coefficients in row $u$ unchanged. No remaining prefix $Q$ can satisfy $c_{u,Q}^2/|Q|\ge T$: if one did, then
		\[
		c_{u,R}+c_{u,Q}\ge\sqrt T\bigl(\sqrt{|R|}+\sqrt{|Q|}\bigr)
		>\sqrt{T(|R|+|Q|)},
		\]
		contradicting the maximal length of $R$. Subsequent edge deletions cannot increase the sum of the largest $i$ coefficients for any $i$. Thus, in the final graph, every prefix of sum greater than $\delta$ has squared sum-of-coefficients divided by its length less than $T$.

		Write the final large coefficients of row $u$ as $c_1\ge c_2\ge\cdots$. Let $E_u$ consist of the vertices in the longest prefix with coefficient sum at most $\delta$. For each rank $i$ outside this prefix,
		\[
		c_i^2\le\frac{1}{i^2}\left(\sum_{h=1}^i c_h\right)^2
		\le\frac{T}{i}.
		\]
		Summing and using $\sum_{i=1}^n1/i\le2\log n$ bounds the squared sum of these large coefficients by $s_A\theta_A d_A$. The small coefficients, including the diagonal if it is below $s_A$, contribute at most
		\[
		\sum_{\substack{v\in M\cap A\\c_{uv}<s_A}}c_{uv}^2
		\le s_A\sum_{w\in N_G(u)}|N_G(w)\cap M\cap A|
		\le s_A\theta_A d_A.
		\]
		By construction, the total coefficient sum over $E_u$ is at most $\delta$, so its marked contribution is also at most $\delta$. This proves~\eqref{eq:coefficients}.

		For each recorded graph, use the graph at the moment of extraction. Its sides are disjoint, its edge count is exactly $c_{u,R}$, and its neighborhood side has at most $\theta_B$ marked vertices. The selection rule gives
		\[
		|R|\le c_{u,R}/s_A,\qquad |N_G(u)|\le d_A,\qquad c_{u,R}>\delta,
		\qquad d_A\theta_B\le\frac{2c_{u,R}^2\theta_B\log n}{s_A\theta_A|R|}.
		\]
		Consequently,
		\[
		(|R|^2|N_G(u)|\theta_B)^{1/3}
		\le(2\log n)^{1/3}c_{u,R} s_A^{-2/3}(\theta_B/\theta_A)^{1/3}.
		\]
		There is at most one recorded graph per row, and the recorded edge lists are disjoint, so $\sum_u c_{u,R}\le m$. The squared bipartite test costs therefore sum to
		\[
		\Ot\left(m/s_A+md_A/\delta
		+m s_A^{-2/3}(\theta_B/\theta_A)^{1/3}\right).
		\]
		Lemma~\ref{lem:or} proves~\eqref{eq:extractcost}. Every deleted edge belongs to a recorded graph, so a collision is detected or survives, except with the assigned test error. Exchanging the sides proves the corresponding statement for $B$. Subsequent deletions only decrease coefficients, preserving both bounds for the same exceptional sets.
	\end{proof}

    \paragraph{Part 4: Estimating common neighbor sums.}

	Now that we have preconditioned the matrix $C_A$, we focus on the vector $C_A\mathbf{1}_{M\cap A}$. This vector intuitively represents a vector indexed by $A$, counting the number of length-two walks to a marked vertex.

	The goal is to estimate a vector $z$ that approximates $C_A\mathbf{1}_{M \cap A}$ in $\ell_{\infty}$-norm. We can do this efficiently now, since \Cref{lem:extract} guarantees that the squared row sum over marked entries outside $E_u$ is small. We can sample uniformly from $M\cap A$ with $\widetilde{O}(\sqrt{n/k})$ queries, where $k=|M\cap A|$. Thus, the $u$th entry of $C_A\mathbf{1}_{M\cap A}$ is $\sum_{v\in M\cap A}c_{uv}=\mathbb{E}[kc_{uv}]$, where the expectation is over a uniform vertex $v\in M\cap A$. The corresponding variance becomes
	\[\underset{v \sim M \cap A}{\mathrm{Var}}[kc_{uv}] \leq k^2 \underset{v \sim M \cap A}{\mathbb{E}} [c_{uv}^2] = k \sum_{v \in M \cap A} c_{uv}^2.\]
	Using concentration, we can thus show that after $t$ samples, with cost $\widetilde{O}(t\sqrt{n/k})$, we have with high probability
	\[\|z - C_A\mathbf{1}_{M \cap A}\|_{\infty} = \widetilde{O}\left(\sqrt{\frac{k}{t} \max_{u \in A} \sum_{v \in M \cap A} c_{uv}^2}\right).\]
    We use this general idea to estimate an approximation of $C_A\mathbf{1}_{M \cap A}$, but we observe from \Cref{lem:extract} that we can't upper bound the entire sum of squared entries of a given row labeled by $u \in A$. Instead, we deal with the exceptional set $E_u$ separately in the proof.
    
	Algorithm~\ref{alg:estimate} makes this idea precise.

	\begin{algorithm}[H]
	\caption{\estimatesums}\label{alg:estimate}
	\small
	\textbf{Input:} A bipartite graph $G$ with vertex sets $A$ and $B$. Let $C_A$ be its common neighbor matrix indexed by $A$. Let $0 < v \leq \delta^2$.
	\begin{steps}
	\item Estimate $r=|M\cap A|$ by $\widehat r$ using Lemma~\ref{lem:quantum}.
	\item If $\widehat r=0$, return the zero vector.
	\item Let $h=\Theta(1+\widehat r v/\delta^2)$.
	\item Obtain $\Theta(\log(n/\epsilon))$ independent batches of $h$ uniform marked samples from $A$.
	\item Compute each batch average $(\widehat r/h)\sum_{i=1}^h c_{uw_i}$ for every row $u$.
	\item Let $z_u$ be the median of these averages, separately for each row.
	\item \textbf{return} $z$.
	\end{steps}
	\end{algorithm}
	
	\begin{lemma}[Sample averages]\label{lem:estimate}
	Let $A$ be a source set of size $a$, write $r=|M\cap A|$, and let $f_u=\sum_{w\in M\cap A}c_{uw}$. Suppose $0<v\le\delta^2$ and that, on a good negative execution, each row has a set $E_u\subseteq A$ such that
	\[
	 \sum_{w\in M\cap E_u}c_{uw}\le\delta,
	 \qquad
	 \sum_{w\in M\cap(A\setminus E_u)}c_{uw}^2\le v.
	\]
	Then Algorithm~\ref{alg:estimate} returns estimates with $|f_u-z_u| \in O(\delta)$ for every row, conditional on a good count estimate and except with the assigned error. The sets $E_u$ are used only in the proof. Conditional on a good count estimate, its query cost, including counting, is
	\begin{equation}\label{eq:estimatecost}
	 \Ot\left(\sqrt a+\sqrt{a|M\cap A|}\,v/\delta^2\right).
	\end{equation}
	\end{lemma}
	\begin{proof}
	First suppose $r=|M\cap A|>0$. Fix a row and split one batch average into the contributions from $E_u$ and $A\setminus E_u$. The latter contribution has variance at most $2\widehat r v/h$. Replacing $r$ by $\widehat r$ in its normalization changes its mean by at most
	\[
	 |\widehat r-r|\sqrt{v/r}=O(\sqrt v)=O(\delta).
	\]
	Chebyshev's inequality therefore bounds its error by $O(\delta)$ with probability at least $7/8$, for a sufficiently large batch constant.
	
	The contribution from $E_u$ is nonnegative and has expectation at most $2\delta$, since $\widehat r/r\le2$. Markov's inequality bounds it by $O(\delta)$ with probability at least $7/8$. Its true contribution is at most $\delta$. Thus the full batch average differs from $f_u$ by $O(\delta)$ with probability at least $3/4$.
	
	Independent batches and a median give failure probability at most $\epsilon/n$ for each row. A union bound gives the claimed simultaneous two-sided accuracy. If $r=0$, a good count estimate is zero and the algorithm correctly returns the zero vector.
	
	Counting costs $\Ot(\sqrt a)$ queries. For $\widehat r>0$, the total sampling limit is $\Ot(\sqrt{a/\widehat r}+\sqrt{a\widehat r}\,v/\delta^2)$. On a good estimate, $\widehat r\le2r$ and $\widehat r\ge1$, proving~\eqref{eq:estimatecost}. If $r=0$, a good estimate returns zero and only the counting cost remains. The sampling limit is fixed once $\widehat r$ is obtained; the bound in terms of the true marked count holds except with the assigned counting error.
	\end{proof}
	
    \paragraph{Total volume reduction subroutine.}
	Algorithm~\ref{alg:reduce} combines the preceding routines with arbitrary sampling, extraction, and threshold parameters. Let $v_A=2s_A\theta_A d_A$ and $v_B=2s_B\theta_B d_B$.

	Algorithm~\ref{alg:cap} stays within its query budget on every input when its counting and sampling calls succeed. Here the budget is guaranteed only on negative inputs satisfying the marked count bounds, so we return \Yes\ on timeout to preserve acceptance on positive inputs.

	\begin{algorithm}[H]
		\caption{\reducevolume}\label{alg:reduce}
		\small
		\textbf{Input:} A bipartite graph $G$ with side sizes $a,b$, degree bounds $d_A,d_B$, marked fraction bound $p\in[0,1]$, and parameters $\theta_A,\theta_B,\delta,\tau_A,\tau_B>0$, $s_A,s_B\ge1$, with $v_A,v_B\le\delta^2$.
		Return \Yes\ if the total query count exceeds a sufficiently large budget $\Ot(Q)$ from Lemma~\ref{lem:implementation}.
		\begin{steps}
			\item $G\gets\samplemarked(G,A,\theta_A)$ and $G\gets\samplemarked(G,B,\theta_B)$.
			\item Run $\prepareestimates$ on $G$ with the supplied parameters, first for rows in $A$ and then with the sides exchanged.
			\item Estimate each side's common neighbor sums by $\estimatesums$, with bounds $(v_A,\delta)$ and $(v_B,\delta)$ respectively.
			\item Minimize $\psi_z(S)$ subject to $|S\cap A|\le ap$ and $|S\cap B|\le bp$.
			\item Let $P=\{u\in A:\ell_u(S)\le\tau_A\}\cup\{u\in B:\ell_u(S)\le\tau_B\}$.
			\item Test the edges outside $G[P]$ with $\covertest$, cover $V(G)\setminus P$, and bound $b_{\mathrm{out}}$.
			\item Return $G[P]$.
		\end{steps}
	\end{algorithm}

	\begin{lemma}[Piece reduction]\label{lem:implementation}
		Let $G$ have $m$ edges and satisfy the input conditions of Algorithm~\ref{alg:reduce}. Suppose $|M\cap A|\le ap$ and $|M\cap B|\le bp$ on the negative inputs under consideration. On a good negative execution the output has marked volume at most
		\[
		a\tau_A+b\tau_B+O(\sqrt{abp\delta}).
		\]
		On the same good negative execution, the edges discarded after estimating the counts have a cover of marked volume at most
		\[
		b_{\mathrm{out}}=c_0p\delta\left(\frac{d_A b}{\tau_A^2}+\frac{d_B a}{\tau_B^2}\right),
		\]
		for a sufficiently large absolute constant $c_0$. The query complexity is $\Ot(Q)$, where
		\begin{align*}
		Q={}&\sqrt{a+b}+\frac{a\sqrt p}{\theta_A}+\frac{b\sqrt p}{\theta_B}
		+Q_{\mathrm{extract}}+\frac{\sqrt p\,(av_A+bv_B)}{\delta^2}+\sqrt{b_{\mathrm{out}}},\\
		Q_{\mathrm{extract}}={}&\sqrt m\left[
		s_A^{-1/3}(\theta_B/\theta_A)^{1/6}
		+s_B^{-1/3}(\theta_A/\theta_B)^{1/6}
		+s_A^{-1/2}+s_B^{-1/2}
		+\sqrt{d_A/\delta}+\sqrt{d_B/\delta}\right].
		\end{align*}
		On a positive input, a collision is detected or survives, except with the assigned error.
	\end{lemma}
	\begin{proof}
		Keep side labels, sizes, and degree bounds fixed during deletions. Lemma~\ref{lem:marked-neighbors} bounds the two sampling costs by
		$\Ot(\sqrt{a+b}+a\sqrt p/\theta_A+b\sqrt p/\theta_B)$ on good count estimates. It establishes the marked-neighbor bounds required by Lemma~\ref{lem:extract}. Applying that lemma on both sides costs $\Ot(Q_{\mathrm{extract}})$ and supplies exceptional contributions at most $\delta$ and squared-sum bounds $v_A,v_B$. The second application preserves the first application's bounds under edge deletions.

		Lemma~\ref{lem:estimate}, using fresh samples from the graph remaining after preprocessing, gives coordinate error $O(\delta)$ at cost
		$\Ot(\sqrt{a+b}+\sqrt p\,(av_A+bv_B)/\delta^2)$.
		Lemma~\ref{lem:two-sides} now gives the retained volume bound and the outside-cover bound $b_{\mathrm{out}}$. Testing this cover costs $\Ot(\sqrt{a+b}+\sqrt{b_{\mathrm{out}}})$ by Lemma~\ref{lem:cover}. Adding the costs proves the claimed bound on good negative executions.

		Choose the global budget large enough to cover these executions. A union bound over the internal error events makes timeout negligible on allowed negative inputs, and the cutoff enforces the worst-case query bound. Every deleted edge is covered by a preceding collision test or neighborhood search. Thus a positive input is detected or retains a collision; returning \Yes\ on timeout preserves this guarantee even when the marked count bounds fail.
	\end{proof}

    \paragraph{Parameter optimization.}
	Fix a target query parameter $t\ge\sqrt n$ and a base cutoff $s\ge1$. At current volume bound $\ell$, partition the edges into $j=O((\log n)^3)$ pieces. For a piece with sides of sizes $a,b$ and $m>t^2$ edges, use the explicit choices
	\begin{equation}\label{eq:iterated-parameters}
	\begin{gathered}
	 p=\min\{1,\gamma \ell/m\},\qquad
	 \tau_A=\frac{\ell}{32ja},\qquad \tau_B=\frac{\ell}{32jb},\qquad
	 \delta=\frac{c\ell t^2}{j^2\gamma^2ab},\\
	 \theta_A=\frac{a\sqrt p}{t},\qquad \theta_B=\frac{b\sqrt p}{t},\qquad
	 s_A=s\sqrt{b/a},\qquad s_B=s\sqrt{a/b},
	\end{gathered}
	\end{equation}
	where $c>0$ is a sufficiently small absolute constant. As before, put $v_A=2s_A\theta_A d_A$ and $v_B=2s_B\theta_B d_B$.

	\begin{algorithm}[H]
		\caption{Collision search with bounded maximum degree}\label{alg:bounded-degree}
		\small
		\textbf{Input:} A known bipartite graph $G$ on at most $n$ labels of maximum degree at most $d$, where $1\le d\le n$.
		\begin{steps}
			\item Let $T=\max\{n^{13/22}d^{3/22},n^{1/2}d^{1/3}\}$ and $t=K\log^{10}(2n)\,T$, for a sufficiently large absolute constant $K$.
			\item If $t\ge\sqrt{nd}$, return $\covertest(G,V(G),nd)$.
			\item Let $s=(nd)^{3/2}/t^3$.
			\item Let $\ell=nd$.
			\item \textbf{while} $\ell>t^2$:
			\item \ind If $G$ is edgeless, return \No.
			\item \ind Partition $G$ using $\partitionedges$; let $j$ be the number of pieces.
			\item \ind Search each piece with at most $t^2$ edges directly; return \Yes\ on a collision, and otherwise discard it.
			\item \ind For every remaining piece, call $\reducevolume$ with the parameters in~\eqref{eq:iterated-parameters} and retain its returned edge list.
			\item \ind Replace $G$ by the union of the retained edge lists and let $\ell\gets \ell/2$.
			\item \textbf{end while}
			\item Return $\covertest(G,V(G),t^2)$.
		\end{steps}
	\end{algorithm}

	\begin{lemma}[Iterated volume reduction]\label{lem:iterated-volume}
		Algorithm~\ref{alg:bounded-degree} decides graph collision with bounded error for bipartite graphs on at most $n$ labels of maximum degree at most $d$, using
		\[
		\Ot\left(\min\left\{\sqrt{nd},
		\max\left\{n^{13/22}d^{3/22},n^{1/2}d^{1/3}\right\}\right\}\right)
		\]
		queries. Its parameter choices satisfy $s_A,s_B\ge1$ and $v_A,v_B\le\delta^2$ in every call to $\reducevolume$.
	\end{lemma}
	\begin{proof}
		Initially the marked volume is at most $nd$. On a good negative execution, each partition piece has marked volume at most $\ell$. By Lemma~\ref{lem:count}, its minimum degrees imply $|M\cap A|\le ap$ and $|M\cap B|\le bp$. The threshold terms sum to $\ell/(16j)$, and
		\[
		\sqrt{abp\delta}\le\frac{\sqrt c\,\ell}{j\sqrt\gamma}\frac{t}{\sqrt m}
		\le\frac{\sqrt c\,\ell}{j\sqrt\gamma}.
		\]
		Thus Lemma~\ref{lem:implementation} retains volume at most $\ell/(4j)$ per piece for sufficiently small $c$. Summing the disjoint retained edge lists gives volume at most $\ell/4$, so halving $\ell$ preserves the invariant. The final cover test rejects on good negative executions. On positive inputs, every deleted collision is covered by a preceding search or test, so a collision is detected or survives. The prescribed reduction cutoffs return \Yes\ and add only their assigned error on negative inputs.

		We bound the costs in Lemma~\ref{lem:implementation}. Sampling costs $\Ot(t)$. Since $\theta_B/\theta_A=b/a$, the first extraction term on either side is at most $\sqrt{nd}\,s^{-1/3}$. Since $m\le\min\{a,b\}\,d$, we have $m\sqrt{a/b}\le\sqrt{m\cdot ad}\le ad\le nd$. Hence $m/s_A\le nd/s$, and symmetrically $m/s_B\le nd/s$. Also,
		\[
		\sqrt{md_A/\delta},\ \sqrt{md_B/\delta}
		\le\Ot\left(\frac{n^{3/2}d}{t\sqrt \ell}\right)
		\le\Ot\left(\frac{n^{3/2}d}{t^2}\right),
		\]
		using $\ell>t^2$, $m\le nd$, and $d_A,d_B\le d$. For estimation, $pd_A\le\gamma^2\ell/a$ gives
		\[
		\frac{a\sqrt p\,v_A}{\delta^2}
		\le\Ot\left(\frac{s(ab)^{5/2}}{\ell t^5}\right)
		\le\Ot\left(\frac{sn^5}{t^7}\right),
		\]
		and the same holds on $B$. Finally, the outside-cover bound is $b_{\mathrm{out}}=O(ct^2)$ because $pm\le\gamma \ell$ and $d_A\le\gamma m/a$, $d_B\le\gamma m/b$, so its test costs $\Ot(t)$. Direct searches and the final cover test also cost $\Ot(t)$ each. Summing over logarithmically many rounds and pieces proves the bound. A union bound with sufficiently small elementary errors gives bounded error overall.

		It remains to verify the parameter choices. If $t\ge\sqrt{nd}$, the direct cover test costs $\Ot(\sqrt{nd})$. This is within a logarithmic factor of the stated minimum. Otherwise $s=(\sqrt{nd}/t)^3\ge1$. A piece with $m>t^2$ has $a,b>t^2/d$ and $a,b\le n$, so
		\[
		s_A,s_B\ge\frac{st}{\sqrt{nd}}=\frac{nd}{t^2}>1.
		\]
		For the variance condition, $ad_A\le\gamma m$, $p\le\gamma \ell/m$, and $d_A\le d$ imply $pd_A^2\le\gamma^2\ell d/a$. Hence
		\[
		\frac{v_A}{\delta^2}
		\le\Ot\left(\frac{s(ab)^{5/2}\sqrt p\,d_A}{\ell^2t^5}\right)
		\le\Ot\left(\frac{s n^{9/2}\sqrt d}{t^8}\right)
		=\Ot\left(\frac{n^6d^2}{t^{11}}\right),
		\]
		using $\ell>t^2$. Since $d\le n$ and $T^{11}\ge n^{13/2}d^{3/2}\ge n^6d^2$, the logarithmic slack in $t$ ensures $v_A\le\delta^2$. The same argument applies to $v_B$.

		The cost calculation above gives
		\[
		\Ot\left(t+\sqrt{nd}\,s^{-1/3}+\sqrt{nd/s}
		+\frac{n^{3/2}d}{t^2}+\frac{sn^5}{t^7}\right).
		\]
		The second term equals $t$, and the third is at most $t$ because $s\ge1$. The fourth is at most $t$ by $T^3\ge n^{3/2}d$. The last equals $n^{13/2}d^{3/2}/t^{10}$ and is at most $t$ by the definition of $T$. Thus the iterative branch costs $\Ot(T)$, proving the claim.
	\end{proof}

    \subsection{Finishing the argument}
    \label{subsec:full-algo}
    
    We can now combine the low- and high-degree algorithms and optimize the degree cut-off $d$ to minimize the resulting query complexity. This gives the final query complexity bound stated in the introduction.

	\begin{proof}[Proof of Theorem~\ref{thm:main}]
		Choose a degree threshold $d$. First apply Lemma~\ref{lem:cap}, which detects a collision or returns a graph of maximum degree at most $d$, preserving a collision if one exists, except with the assigned error. This costs $\Ot(n^{3/4}/d^{1/4})$ queries. Apply Lemma~\ref{lem:double} to the returned graph; this gives a bipartite graph on $2n$ labels whose maximum degree is still at most $d$. Apply Lemma~\ref{lem:iterated-volume} to this bipartite graph. Its cost is at most
		\[
		\Ot\left(\max\left\{n^{13/22}d^{3/22},n^{1/2}d^{1/3}\right\}\right).
		\]
		Balancing the high-degree cost with the first low-degree term gives
		\[
		\frac{n^{3/4}}{d^{1/4}}=n^{13/22}d^{3/22},
		\qquad d=\lceil n^{7/17}\rceil.
		\]
		Both balanced terms are $\Ot(n^{11/17})$. The remaining term is smaller:
		\[
		n^{1/2}d^{1/3}=O(n^{65/102})=O(n^{11/17}).
		\]
		Thus the total query complexity is $\Ot(n^{11/17})$. On a negative input both routines reject, and on a positive input the first detects a collision or preserves one for the second. Amplifying each routine to error at most $1/6$ gives total error at most $1/3$. For bounded $n$, query all input bits. Repeating the algorithm $O(\log n)$ times and taking the majority gives the stated high probability guarantee.
	\end{proof}
	
	\section{Functions with positive certificates of size two}\label{sec:certificates}
	The preceding algorithm also applies to functions with positive certificates of size at most two over an arbitrary finite alphabet.
	
	Let $\Sigma$ be an alphabet of size $q$, and let $f:D\to\{0,1\}$, where $D\subseteq\Sigma^n$. A positive certificate is a partial assignment that forces the value $1$ among inputs in $D$. Let $C_1(f)$ be the least nonnegative integer such that every positive input has a positive certificate of size at most $C_1(f)$. Write $Q_2(f)$ for the quantum query complexity with error at most $1/3$.
	
	\begin{lemma}[Reduction to graph collision]\label{lem:certificate-reduction}
		Let $f:D\to\{0,1\}$ be a (partial) function, where $D\subseteq\Sigma^n$, $n\ge2$, $|\Sigma|=q$, and $C_1(f)\le2$.\footnote{If $n=1$, one query evaluates $f$. For binary inputs, the comparison with graph collision follows by fixing one endpoint of a single edge to be marked.} There is a known simple graph $G$ on $nq$ vertices and a bit string $y(x)$ such that, for every $x\in D$,
		\[
		f(x)=\operatorname{gc}_G(y(x)),\qquad
		\operatorname{gc}_G(y)=\bigvee_{\{u,v\}\in E(G)}(y_u\wedge y_v).
		\]
		Each query to $y(x)$ can be simulated with at most two queries to $x$. Consequently, $Q_2(f)\le2Q_2(\operatorname{gc}_G)$. For binary inputs, one query suffices and the factor $2$ can be omitted.
	\end{lemma}
	\begin{proof}
		Suppose first that $f$ is nonconstant on $D$. Let $V(G)=[n]\times\Sigma$ and $y_{(i,a)}(x)=\mathbf1[x_i=a]$. Thus the vertex $(i,a)$ is marked exactly when $x_i=a$.

		For each positive certificate $\{(i,a),(j,b)\}$ of size two, include the edge $\{(i,a),(j,b)\}$ in $E(G)$. For each singleton positive certificate $\{(i,a)\}$, choose a coordinate $j\ne i$ and include the edges $\{(i,a),(j,b)\}$ for all $b\in\Sigma$, adding each edge only once throughout the construction.

		An edge arising from a certificate of size two has both endpoints marked exactly when $x$ satisfies that certificate. For a singleton certificate $\{(i,a)\}$, exactly one vertex $(j,b)$ is marked, so one of the added edges has both endpoints marked exactly when $x_i=a$. Thus $G$ has a marked edge if and only if $x$ satisfies a positive certificate of size at most two. Since $C_1(f)\le2$, this is equivalent to $f(x)=1$ for every $x\in D$.
		
		For constant value $0$, use the same input bits on an edgeless graph. For constant value $1$, include every edge between $\{1\}\times\Sigma$ and $\{2\}\times\Sigma$. Exactly one vertex in each part is marked, so a collision always exists.
		
		To simulate a query to $y_{(i,a)}$, query $x_i$ into a work register, flip the answer bit if $x_i=a$, and uncompute the work register. This uses two input queries and implements the oracle for $y$.
		
		For binary inputs, $y_{(i,a)}(x)=x_i\oplus(1\oplus a)$. Query $x_i$ directly into the answer bit and flip that bit when $a=0$. This uses one query.
	\end{proof}
	
	\begin{theorem}[Query bound for an arbitrary alphabet]\label{thm:certificate-bound}
		Let $f:D\to\{0,1\}$ be a (partial) function, where $D\subseteq\Sigma^n$, $|\Sigma|=q$, and $C_1(f)\le2$. Then $Q_2(f)=\Ot\bigl((nq)^{11/17}\bigr)$.
	\end{theorem}
	\begin{proof}
		Apply Lemma~\ref{lem:certificate-reduction} and Theorem~\ref{thm:main} to the graph on $nq$ vertices. Two queries to $x$ per query to $y$ give the stated bound.
	\end{proof}
	
	\begin{corollary}[Bounds for different alphabet sizes]\label{cor:alphabet-bounds}\label{cor:constant-alphabet}\label{cor:subpolynomial-alphabet}
		For every function $f$ as in Theorem~\ref{thm:certificate-bound}:
		\begin{enumerate}[label=(\roman*)]
			\item If $q=O(1)$, then $Q_2(f)=\Ot(n^{11/17})$.
			\item If $q=n^{o(1)}$, then $Q_2(f)\le n^{11/17+o(1)}=o(n^{2/3})$.
			\item If $q=O(n^{1/33-\varepsilon})$ for some fixed $\varepsilon>0$, then $Q_2(f)=o(n^{2/3})$.
		\end{enumerate}
	\end{corollary}
	\begin{proof}
		Substitute the three bounds on $q$ into Theorem~\ref{thm:certificate-bound}. In the third case, $(nq)^{11/17}=O(n^{2/3-11\varepsilon/17})$, so the logarithmic factors do not affect the strict improvement.
	\end{proof}
	
	\begin{corollary}[Equivalence for binary inputs]\label{cor:certificate-equivalence}
		Let
		\[
		q_{\rm cert}(n)=
		\max_{\substack{f:\{0,1\}^n\to\{0,1\}\\C_1(f)\le2}}Q_2(f),
		\qquad
		q_{\rm coll}(n)=\max_{|V(G)|=n}Q_2(\operatorname{gc}_G).
		\]
		Then $q_{\rm coll}(n)\le q_{\rm cert}(n)\le q_{\rm coll}(2n)$.
	\end{corollary}
	\begin{proof}
		Every graph collision function has positive certificate complexity at most two, since the two marked endpoints of an edge certify a positive input. This gives the first inequality. The binary case of Lemma~\ref{lem:certificate-reduction} gives the second.
	\end{proof}

	\section{A lower bound breaking the \texorpdfstring{$\Omega(\sqrt n)$}{Omega(sqrt(n))} barrier}\label{sec:lower-bound}
	We now give a family of graphs for which graph collision requires $\Omega(\sqrt{n\log n})$ queries. The graphs are incidence graphs of points and lines over a finite field. The proof uses the fact that two points determine at most one line and two lines meet in at most one point.

	Let $q$ be a prime power. Let $P$ and $L$ be disjoint copies of $\mathbb F_q^2$, representing points and nonvertical lines, respectively. A point $(a,b)$ is incident to a line $(s,t)$ when $b=sa+t$. Let $G_q$ be the Levi graph associated to this incidence structure, i.e., the bipartite graph with one independent set of vertices corresponding to the set of points in $P$ and the other independent set of vertices corresponding to the set of lines in $L$. Place an edge $(u,v)$ in $E(G_q)$ if the point in $P$ corresponding to $u$ is contained in the line in $L$ corresponding to $v$. Thus, the graph $G_q$ has $2q^2$ vertices, degree $q$, and $q^3$ edges.

	\begin{theorem}\label{thm:incidence-lower-bound}
		For all sufficiently large prime powers $q$, we have $Q_2(\operatorname{gc}_{G_q})=\Omega(q\sqrt{\log q})$. Consequently, for sufficiently large values of $n$, there is an $n$-vertex graph $G$ with $Q_2(\operatorname{gc}_G)=\Omega(\sqrt{n\log n})$.
	\end{theorem}

	By Lemma~\ref{lem:lower-adversary}, it suffices to construct a matrix $\Gamma$ with rows indexed by positive inputs and columns indexed by negative inputs, such that $\|\Gamma\|/\max_v\|\Gamma\circ\Delta_v\|=\Omega(q\sqrt{\log q})$. The matrix $\Gamma\circ\Delta_v$ keeps only entries whose two inputs disagree at the queried vertex $v$. We want the norm of the full matrix to be large compared to the norm of each such restriction. 

	\subsection{Construction of the adversary matrix}\label{subsec:lower-construction}
	We compare inputs with no marked edge to inputs with exactly one marked edge $a\ell$, where $a \in P$ and $\ell \in L$. We first specify distributions on these inputs.

	\paragraph{Input distributions.}
	Let
	\[
	 r=\frac{\ln q}{8q},\qquad p=r(1-r)^{-(q-1)},\qquad
	 \bar r=1-r,\qquad \bar p=1-p.
	\]
	Here $r$ is the probability of marking a line, and $p$ is the probability of marking an uncovered point. We say a point is uncovered if it is not contained in any marked line. We have $p=q^{-7/8+o(1)}$. For sufficiently large $q$,
	\begin{equation}\label{eq:lower-probabilities}
	 4r\le p<\frac14,\qquad qp\ge1,\qquad qp^2\le1.
	\end{equation}
	For an input $x$, let $\operatorname{free}_u(x)=\prod_{h\in N(u)}(1-x_h)$. Thus $u$ is uncovered exactly when $\operatorname{free}_u(x)=1$.

	For the negative distribution $\mu^-$, mark the lines independently with probability $r$, and then mark each uncovered point independently with probability $p$. For an edge $a\ell$, define $\mu^{a\ell}$ by fixing $x_a=x_\ell=1$, fixing the other points on $\ell$ and the other lines through $a$ to zero, and using the same sampling rule on the remaining vertices. The only marked edge is $a\ell$. In particular, the positive distributions have disjoint supports.

	The negative sampling rule prevents a collision at every point: a marked line forces all its incident points to zero. The positive distribution can also be obtained by conditioning a negative sample on $x_a=1$, marking $\ell$, and setting the other points on $\ell$ to zero. All lines through $a$ were already unmarked, and all other bits remain unchanged. Thus the two distributions are related by a change confined to one line and its incident points.
	Figure~\ref{fig:hard-distributions} illustrates the two distributions.

	\paragraph{Choice of probabilities.}
    The aim is to have many marked points and lines among which to hide the planted edge. The two sides compete: marking more lines leaves fewer points available to be marked. Each point lies on $q$ lines, so it has $qr$ marked incident lines on average and is uncovered with probability $(1-r)^q$.

    We compensate for this loss by marking uncovered points with the larger probability $p$. The relation $p(1-r)^{q-1}=r$ makes the probability that a given point is marked equal to $p(1-r)^q=r(1-r)$, almost the same as the line probability $r$. Thus the two sides have nearly the same density of marked vertices. If $r$ is of order $1/q$, each side has only order $q$ marked vertices on average. With $r=\ln q/(8q)$, each side instead has order $q\ln q$ marked vertices on average, while the sampling rule still prevents every collision on a negative input.

    Why not mark even more lines? The probability of being uncovered decreases exponentially in $qr$, so the compensating probability $p=r(1-r)^{-(q-1)}$ grows rapidly. In particular, $p\ge r e^{(q-1)r}$, and taking $r$ asymptotically larger than $\ln q/q$ would force $p>1$. The logarithmic scale therefore allows many more marked vertices while preserving this balance. The small factor $1/8$ keeps the conditional point probability small as well: $(1-r)^q=q^{-1/8+o(1)}$ and $p=q^{-7/8+o(1)}$.

    \paragraph{Bases indexed by independent sets.}

	The input bits are dependent, since a marked line forces its neighboring points to zero. Under $\mu^-$, once the line bits are fixed, the uncovered point bits are independent with mean $p$, and every covered point bit is zero. Thus the conditional expectation of $x_u$ is $p \cdot \operatorname{free}_u(x)$. Subtracting this expectation gives a function with conditional mean zero; this subtraction is called centering. Dividing by $\sqrt{p\bar p}$ then gives conditional second moment one at uncovered points. For a line bit, we similarly subtract its mean $r$ and divide by $\sqrt{r\bar r}$.
	In each distribution space, use the inner product $\langle f,g\rangle=\mathbb E[\overline f g]$. Let
	\[
	 \xi_u=\frac{x_u-p\operatorname{free}_u(x)}{\sqrt{p\bar p}},
	 \qquad \eta_h=\frac{x_h-r}{\sqrt{r\bar r}},
	 \qquad \xi_S=\prod_{u\in S}\xi_u,\qquad
	 \eta_T=\prod_{h\in T}\eta_h.
	\]
	As usual, $N(S)=\bigcup_{u\in S}N(u)$. Define functions
	\begin{align*}
	 \phi^-_{S,T}&=\bar r^{-|N(S)|/2}\xi_S\eta_T,
	 &S&\subseteq P, &T&\subseteq L\setminus N(S),\\
	 \phi^{a\ell}_{S,T}&=\bar r^{-|N(S)\setminus N(a)|/2}\xi_S\eta_T,
	 &S&\subseteq P\setminus\ell, &T&\subseteq L\setminus(N(a)\cup N(S)).
	\end{align*}
	Here $P\setminus\ell$ excludes the points incident to $\ell$. The sets $S$ and $T$ index basis vectors; they are not the marked sets of an input.

	Each basis vector is a function of the input $x$. The product $\xi_S$ vanishes unless every line in $N(S)$ is unmarked. On those inputs, the functions $\eta_h$ for $h\in N(S)$ are constant, so only lines outside $N(S)$ need separate indices. This explains why the basis is indexed by independent sets. The normalizing factors account for the probability that the required lines are unmarked.

	For example, $\phi^-_{\{u\},\varnothing}=\bar r^{-q/2}\xi_u$ is zero when a marked line covers $u$. When $u$ is uncovered, $\xi_u=(x_u-p)/\sqrt{p\bar p}$. Adding a line $h\notin N(u)$ to the index set multiplies this function by $\eta_h$. A line $h\in N(u)$ supplies no additional factor that varies on its support: there $x_h=0$, so $\eta_h$ is constant. The indices therefore specify which functions $\xi_u$ and $\eta_h$ occur in the product.

	\begin{lemma}[Independent set bases]\label{lem:lower-bases}
		The functions $\phi^-_{S,T}$ form an orthonormal basis for functions on the support of $\mu^-$, and the functions $\phi^{a\ell}_{S,T}$ form an orthonormal basis for functions on the support of $\mu^{a\ell}$, with the inner products above.
	\end{lemma}
	\begin{proof}
		Condition on the line bits. The functions $\xi_u$ occurring in the products are independent and have mean zero, and $\mathbb E[\xi_u^2\mid(x_h)_{h\in L}]=\operatorname{free}_u(x)$. Distinct point sets are therefore orthogonal. For equal point sets $S$, the second moment forces every line in $N(S)$ to be zero. The remaining line bits are independent, so the products of their functions $\eta_h$ are orthonormal. The displayed powers of $\bar r$ normalize the squared norms, taking account of the lines through $a$ that are already fixed in the positive distribution.

		For the negative distribution, both the basis indices and the supported inputs are the independent sets of $G_q$. For the positive distribution, both are the independent sets of the graph remaining after deleting the points on $\ell$ and the lines through $a$. Their cardinalities agree, so the orthonormal families are complete.
	\end{proof}

	Let $U^-$ and $U^+$ be the matrices taking these basis coordinates to input coordinates. Their entries are
	\[
	 U^-_{x,(S,T)}=\sqrt{\mu^-(x)}\,\phi^-_{S,T}(x),
	 \qquad
	 U^+_{y,(a,\ell,S,T)}=\sqrt{\mu^{a\ell}(y)}\,\phi^{a\ell}_{S,T}(y),
	\]
	where the latter entry is zero unless $y$ has unique marked edge $a\ell$. These matrices are orthogonal by Lemma~\ref{lem:lower-bases} and the disjointness of the positive supports.

	Write $\Pi_v^-$ for multiplication by $x_v$ in the negative basis, and $\Pi_v^+$ for multiplication by $x_v$ in the direct sum of the positive bases. These are orthogonal projections onto functions supported on inputs whose bit at $v$ is one. 

	\paragraph{The planting map.}
	To obtain an input with marked edge $a\ell$, start with a negative input whose point $a$ is marked, mark $\ell$, and unmark the other points on $\ell$. Every line through $a$ is initially unmarked, so this operation creates exactly one marked edge. We construct matrix $B$ by expressing this operation in the bases above.

	The condition that $a$ is marked has two contributions in the centered basis. Indeed, the definition of $\xi_a$ gives
	\[
	 \frac{x_a}{\sqrt p}
	 =\sqrt p\,\operatorname{free}_a(x)+\sqrt{\bar p}\,\xi_a(x).
	\]
	The first term requires all lines through $a$ to be unmarked, and the second contains the point index $a$. This explains the two terms and their coefficients in $B$.

	For a set $F$ of lines, let $z_F$ be the unit vector indexed by $J\subseteq F$ with coordinates $z_F(J)=(-\sqrt r)^{|J|}\bar r^{(|F|-|J|)/2}$. The binomial theorem gives its unit norm, and expanding the product gives
	\begin{equation}\label{eq:lower-zero-lines}
	 \sum_{J\subseteq F}z_F(J)\eta_J
	 =\bar r^{-|F|/2}\prod_{h\in F}(1-x_h).
	\end{equation}
	Thus $z_F$ represents the condition that all lines in $F$ are unmarked.

	For a positive row $(a,\ell,S,T)$, the lines in $N(a)\cap N(S)$ are already unmarked wherever $\xi_S$ is nonzero. Thus only the lines in $F=N(a)\setminus N(S)$ need to be included in $z_F$. In particular, $\ell\in F$. Every point in $S$ covers at most one line through $a$, so
	\begin{equation}\label{eq:lower-covered-lines}
	 |F|\ge q-|S|.
	\end{equation}
	With the normalizations of our bases, the two contributions above give the planting map
	\[
	 (Bf)_{a,\ell,S,T}
	 =\sqrt p\sum_{J\subseteq F}z_F(J)f_{S,T\cup J}
	 +\sqrt{\bar p}\,f_{S\cup\{a\},T},
	\]
	for a vector $f$ in the negative basis. The first term removes a subset of lines through $a$, and the second removes the point index $a$. The signs come from \eqref{eq:lower-zero-lines}, and $|z_F(J)|^2$ is the probability of selecting exactly $J$ when each line in $F$ is selected independently with probability $r$.

	\begin{lemma}[Planting an edge]\label{lem:lower-planting}
		On a negative input with $x_a=1$, let $\operatorname{plant}_{a\ell}$ mark $\ell$ and unmark every other point on $\ell$. If $B^{a\ell}$ is the block of $B$ associated with $a\ell$, then, for every function $f$ on positive inputs,
		\[
		 ((B^{a\ell})^*f)(x)=
		 \begin{cases}
		 (r\bar r)^{-1/2}f(\operatorname{plant}_{a\ell}(x)),&x_a=1,\\
		 0,&x_a=0.
		 \end{cases}
		\]
		Consequently, $\Pi_v^+B^{a\ell}=B^{a\ell}\Pi_v^-$ for every bit preserved by this operation, and $B^{a\ell}\Pi_\ell^-=0$.
	\end{lemma}
	\begin{proof}
		Expanding the subset sum using \eqref{eq:lower-zero-lines} gives
		\[
		 (B^{a\ell})^*\phi^{a\ell}_{S,T}
		 =\bar r^{-|N(S\cup\{a\})|/2}\xi_S\eta_T\,
		   \frac{x_a}{\sqrt p}.
		\]
		Now $|N(S\cup\{a\})|=q+|N(S)\setminus N(a)|$ and $p\bar r^q=r\bar r$. Planting changes only $\ell$ and its incident points. Since $S$ contains no point on $\ell$ and $\ell\notin T$, the factors $\xi_S\eta_T$ are unchanged. This proves the formula on every basis vector. Preserved bits give the projection identities, and $x_ax_\ell=0$ on negative inputs gives the last assertion.
	\end{proof}

	\paragraph{The additive correction.}
	Planting changes the input bit at $\ell$ from zero to one. A query to $\ell$ therefore distinguishes each input from its planted image. We add a map $C$ to cancel part of this contribution. Signed entries are allowed in the general adversary bound.

	Consider first $S=\varnothing$ and the source coordinates without the point index $a$. On the two line coordinates $1,\eta_\ell$, the normalized condition $x_\ell=0$ is represented by $(\sqrt{\bar r},-\sqrt r)$. The vector $z_F$ is a product of these unit vectors. Taking the inner product in the $\ell$ coordinate therefore removes that coordinate from the subset term of $B$. Its coefficient at the empty subset of $F\setminus\{\ell\}$ becomes $\sqrt p\,\bar r^{(q-1)/2}=\sqrt r$. In contrast, the coordinate containing just the line index $\ell$ has inner product $-\sqrt r$ with the same condition. This suggests removing the line index with coefficient one. Define
	\[
	 (Cf)_{a,\ell,S,T}=f_{S,T\cup\{\ell\}}.
	\]
	The two empty subset coefficients now cancel. Every other subset contribution contains an additional line through $a$. Knowing such a line restricts the possible choices of $a$ to its points. For nonempty $S$, the same calculation leaves the coefficient $\sqrt p\,\bar r^{(|F|-1)/2}-\sqrt r$. Its difference from zero comes only from the lines in $N(a)\cap N(S)$, which are already forced to be unmarked. There are at most $|S|$ such lines, since each point of $S$ lies on at most one line through $a$.

	\paragraph{The multiplicative correction.}

	We implement the multiplicative correction by weighting the columns. A source coordinate with many point and line indices offers more choices for the planted edge: a point removed by $B$ specifies $a$, and a line removed by $C$ specifies $\ell$. We give smaller weights to columns with more indices to limit their combined contribution. At the same time, we keep weight one at the coordinate indexed by $S=T=\varnothing$, which contributes to every planted edge.
	Let
	\[
	 k=\left\lceil\frac1{qp^2}\right\rceil,
	 \qquad d_{S,T}=\frac{k}{k+|S|+|T|}.
	\]
	These weights decrease gradually as indices are added: they are at least $1/2$ when $|S|+|T|\le k$, and decrease in proportion to $k/(|S|+|T|)$ above this scale. They also have a direct interpretation. Uniformly order the $|S|+|T|$ indices together with $k$ auxiliary elements. The weight $d_{S,T}$ is the probability that the first element is auxiliary. Let $D$ be diagonal with these entries, and define
	\begin{equation}\label{eq:lower-construction}
	 \widetilde{\Gamma}=(B+C)D,\qquad
	 \Gamma=U^+\widetilde{\Gamma}(U^-)^*.
	\end{equation}
	The matrix $\widetilde{\Gamma}$ is written in the point and line bases, while $\Gamma$ is written in input coordinates. All three operations from a source coordinate receive the same weight, and point and line indices are treated equally. We have $\|D\|\le1$, and the source coordinate indexed by \mbox{$S=T=\varnothing$} has weight one. The rows of $\Gamma$ are indexed by positive inputs and its columns by negative inputs, as in Lemma~\ref{lem:lower-adversary}.

	Both corrections are designed to control the denominator of the adversary ratio. The additive correction $C$ creates cancellations, and the multiplicative correction $D$ reduces the weights of columns with many indices. Both preserve the image of the coordinate indexed by $S=T=\varnothing$. Let $e_\varnothing$ be the corresponding unit vector. We have $Ce_\varnothing=0$ and $De_\varnothing=e_\varnothing$, so
	\[
	 \widetilde{\Gamma}e_\varnothing=(B+C)De_\varnothing=Be_\varnothing.
	\]
	Since $U^+$ and $U^-$ are orthogonal, this gives $\|\Gamma\|=\|\widetilde{\Gamma}\|\ge\|Be_\varnothing\|$. Thus the corrections preserve this lower bound on the numerator.

	\paragraph{The numerator.}
	We evaluate $Be_\varnothing$ to obtain the lower bound on the numerator.

	Each possible planted edge contributes to a different positive coordinate. These contributions are orthogonal, so their squared magnitudes add over all $q^3$ edges. The choice of $r$ makes this sum of order $q^2\log q$, which explains the additional factor $\sqrt{\log q}$ in the lower bound.

	\begin{lemma}[Numerator bound]\label{lem:lower-numerator}
		The matrices in \eqref{eq:lower-construction} satisfy $\|\Gamma\|=\|\widetilde{\Gamma}\|\ge\sqrt{q^3r\bar r}=\Omega(q\sqrt{\log q})$.
	\end{lemma}
	\begin{proof}
		We have $\|\Gamma\|=\|\widetilde{\Gamma}\|\ge\|Be_\varnothing\|$. For each planted edge $a\ell$, the vector $Be_\varnothing$ has coefficient $\sqrt p\,\bar r^{q/2}=\sqrt{r\bar r}$ in the positive coordinate with \mbox{$S=T=\varnothing$}. These $q^3$ coordinates are orthogonal, and their squared coefficients sum to $q^3r\bar r$, proving the claim.
	\end{proof}

	\subsection{The denominator and the adversary ratio}\label{subsec:lower-denominator}

	The denominator measures how much of the adversary matrix connects inputs that differ at one queried vertex. A map that preserves this input bit gives the same result whether we restrict to inputs with that bit equal to one before or after applying the map. The corresponding difference between the two operators is therefore zero.
	We bound $\|\Gamma\circ\Delta_v\|$ by calculating in the bases above. In input coordinates, the matrix with entries $(y_v-x_v)\Gamma_{y,x}$ becomes $\Gamma\circ\Delta_v$ after multiplying column $x$ by $1-2x_v$. These sign changes preserve the norm. Changing to the point and line bases therefore gives
	\begin{equation}\label{eq:lower-disagreement}
	 \|\Gamma\circ\Delta_v\|
	 =\|\Pi_v^+\widetilde{\Gamma}-\widetilde{\Gamma}\Pi_v^-\|.
	\end{equation}
	There are two contributions to this difference. The first comes from applying the query projection before or after $B+C$. The second comes from the column weights: the query projection can mix basis coordinates with different values of $|S|+|T|$, and hence different weights. Let $E_v=\Pi_v^+(B+C)-(B+C)\Pi_v^-$. Adding and subtracting $(B+C)\Pi_v^-D$ gives
	\[
	 \Pi_v^+\widetilde{\Gamma}-\widetilde{\Gamma}\Pi_v^-
	 =E_vD+(B+C)(\Pi_v^-D-D\Pi_v^-).
	\]

	We will bound the first contribution by counting. For the second, the reciprocal form of the weights reduces the calculation to the change in the number of indices. Let $\mathsf N$ be diagonal with entries $|S|+|T|$. Since $D^{-1}=I+\mathsf N/k$, expanding the right side gives the identity
	\[
	 \Pi_v^-D-D\Pi_v^-=-\frac1kD(\Pi_v^-\mathsf N-\mathsf N\Pi_v^-)D.
	\]
	Using \eqref{eq:lower-disagreement}, $\|\widetilde{\Gamma}\|=\|\Gamma\|$, and $\|D\|\le1$, we obtain
	\begin{equation}\label{eq:lower-error-split}
	 \frac{\|\Gamma\circ\Delta_v\|}{\|\Gamma\|}
	 \le\frac{\|E_vD\|}{\|\Gamma\|}
	 +\frac{\|\Pi_v^-\mathsf N-\mathsf N\Pi_v^-\|}{k}.
	\end{equation}
	The factor $(B+C)D=\widetilde{\Gamma}$ in the second contribution has norm $\|\Gamma\|$ and disappears when we divide by this norm. Thus we only need to control how the query changes the number of indices; no separate bound on $\|B+C\|$ is required.

	We will show that $\|E_vD\|=O(1)$ and $\|\Pi_v^-\mathsf N-\mathsf N\Pi_v^-\|=O(\sqrt p(1+qr))$. Together with Lemma~\ref{lem:lower-numerator}, these estimates give the required adversary ratio. We first describe the local query projections and bound the change in the number of indices. We then bound $E_v$ by counting and apply the column weights.

	\paragraph{The action of a query.}
	A query to a free line acts on the two coordinates in which that line index is absent or present. A point query has a slightly larger block. When the point index $u$ is present, the independent set condition excludes every line through $u$. When $u$ is absent, the free lines through $u$ can occur in any subset. Fixing all other indices groups these possibilities into one block. For one line, write $z=(\sqrt{\bar r},-\sqrt r)^{\mathsf T}$. The following lemma gives the query projections on these blocks.

	\begin{lemma}[Local query projections]\label{lem:lower-queries}
		On the two coordinates of a free line, its query projection is
		\[
		 I-zz^*=\begin{pmatrix}r&\sqrt{r\bar r}\\\sqrt{r\bar r}&\bar r\end{pmatrix}.
		\]
		For a point $u\notin S$, fix the line indices outside $N(S)\cup N(u)$ and let $F=N(u)\setminus N(S)$. The negative point projection on the coordinates $(S,T\cup J)$, $J\subseteq F$, and $(S\cup\{u\},T)$ is
		\begin{equation}\label{eq:lower-point-query}
		 \Pi_u^-=m_Fm_F^*,\qquad m_F=\sqrt p\,z_F\oplus\sqrt{\bar p}.
		\end{equation}
		In a positive space, the same formula holds after deleting the lines in $N(a)$. The projections at $a$ and $\ell$ are the identity; those at the other points on $\ell$ and the other lines through $a$ are zero.
	\end{lemma}
	\begin{proof}
		The line formula follows by multiplying the functions $1,\eta_h$ by $x_h$. For a point, use \eqref{eq:lower-zero-lines} and the identities
		\[
		 x_u=p\operatorname{free}_u+\sqrt{p\bar p}\,\xi_u,
		 \qquad x_u\xi_u=\bar p\xi_u+\sqrt{p\bar p}\operatorname{free}_u.
		\]
		Order the subset coordinates first and the point coordinate last. By \eqref{eq:lower-zero-lines}, multiplication by $\operatorname{free}_u$ on the subset coordinates is the projection $z_Fz_F^*$. The two identities above, with the normalizations in the basis definitions, give the blocks
		\[
		 \Pi_u^-=
		 \begin{pmatrix}
		 pz_Fz_F^* & \sqrt{p\bar p}\,z_F\\
		 \sqrt{p\bar p}\,z_F^* & \bar p
		 \end{pmatrix}
		 =m_Fm_F^*.
		\]
		The off-diagonal blocks exchange the point coordinate with the combination $z_F$ of subset coordinates. Since $\|z_F\|=1$, we have $\|m_F\|^2=p+\bar p=1$, so this is a rank one orthogonal projection. The positive formulas follow from the fixed input bits.
	\end{proof}

	\paragraph{The number of indices changed by a query.}
	The second term of \eqref{eq:lower-error-split} depends only on the difference between the sizes occurring in a local query block. A point query compares a subset of free lines with a single point index.

	The squared coefficients of $z_F$ describe independent line selections with probability $r$. Thus the size change is controlled by the second moment of the number of selected lines. The coefficient $\sqrt p$ in the point query vector supplies the additional small factor in the bound.

	\begin{lemma}[Change in the number of indices]\label{lem:lower-line-count}
		For every vertex $v$,
		\[
		 \|\Pi_v^-\mathsf N-\mathsf N\Pi_v^-\|=O\bigl(\sqrt p(1+qr)\bigr).
		\]
	\end{lemma}
	\begin{proof}
		Consider a point query block from \eqref{eq:lower-point-query}, with base point set $S$ and external line set $T$. Subtract $(|S|+|T|+1)I$ from $\mathsf N$, which leaves its commutator with the query projection unchanged. The shifted diagonal entries are $|J|-1$ on subset coordinates and zero on the point coordinate. Hence
		\[
		 \bigl\|(\mathsf N-(|S|+|T|+1)I)m_F\bigr\|^2
		 =p\,\mathbb E[(j-1)^2],
		 \qquad j\sim\operatorname{Bin}(|F|,r).
		\]
		Counting selected lines and ordered pairs gives $\mathbb E[(j-1)^2]=1-|F|r+|F|(|F|-1)r^2\le(1+qr)^2$. Expanding the difference with the rank one projection $m_Fm_F^*$ as two rank one maps bounds its norm by twice the displayed vector norm.

		A free line query changes the total size by one. On its two coordinates, the commutator block is
		\[
		 \Pi_h^-\mathsf N-\mathsf N\Pi_h^-
		 =\begin{pmatrix}0&\sqrt{r\bar r}\\-\sqrt{r\bar r}&0\end{pmatrix},
		\]
		whose norm is $\sqrt{r\bar r}\le\sqrt p$. A line fixed to zero gives no contribution. The local blocks are orthogonal, so their largest norm proves the claim.
	\end{proof}

	\Needspace{12\baselineskip}
	\paragraph{Counting the contributions.}
	We now turn to the first term of \eqref{eq:lower-error-split} and bound $E_v$ by counting compatible planted edges. For each type of contribution, we bound the norm of a local map and then count how many such maps can use a fixed source coordinate. Orthogonal target spaces allow us to add squared norms. Thus a source used by at most $t$ maps of norm at most $\sqrt p$ receives a squared bound of at most $pt$. The following observation makes this calculation valid for arbitrary vectors, even when different maps use some of the same source coordinates.

	\begin{lemma}[Weighted counting]\label{lem:lower-counting}
		Suppose maps $D_j$ have orthogonal target spaces, each $D_j$ reads only source coordinates in $I_j$, and $\|D_j\|\le w_j$. Then
		\[
		 \left\|\sum_jD_jf\right\|^2
		 \le\sum_i\left(\sum_{j:i\in I_j}w_j^2\right)|f_i|^2.
		\]
	\end{lemma}
	\begin{proof}
		Let $Q_j$ project onto $I_j$. Since $D_j=D_jQ_j$, we have $\|D_jf\|^2\le w_j^2\|Q_jf\|^2$. Orthogonality of the target spaces gives
		\[
		 \left\|\sum_jD_jf\right\|^2
		 =\sum_j\|D_jf\|^2
		 \le\sum_jw_j^2\sum_{i\in I_j}|f_i|^2.
		\]
		Interchanging the two sums gives the claimed bound. Each source coordinate is counted once for every local map that can use it.
	\end{proof}

	For a fixed planted edge, fixing the point indices and the line indices outside the queried neighborhoods gives the local blocks to which this lemma applies. A point query groups together the coordinates with its point index absent and present. Distinct planted edges have orthogonal target spaces. We split the matrix into a constant number of types of contributions; combining their bounds costs only a constant factor.

	We next explain the cancellation supplied by $C$. For a query to the planted line $\ell$, the positive bit is fixed to one, so the difference on this block is $(B+C)(I-\Pi_\ell^-)$. In the subset and line removal terms, $\ell$ is a free line coordinate. On these coordinates, $I-\Pi_\ell^-=zz^*$, so we first multiply the corresponding part of the row by $z$. Suppose $S=\varnothing$. Among the terms with no other line through $a$, the subset term of $B$ contributes $\sqrt p\,\bar r^{(q-1)/2}$, and the line removal term $C$ contributes $-\sqrt r$. These are equal in magnitude by our choice of $p$, so they cancel.

	The terms left after this cancellation contain another line through $a$, whose presence will restrict the possible planted edges. For nonempty $S$, the lines in $N(a)\cap N(S)$ must already be unmarked wherever $\xi_S$ is nonzero. This changes the empty subset coefficient, but each point in $S$ can account for at most one such line. The next lemma separates the nonempty subsets from this remaining coefficient. It also gives the corresponding formula for a query to another point on the planted line.

	\begin{lemma}[Queries on the planted line]\label{lem:lower-cancellation}
		Fix a positive row $(a,\ell,S,T)$, and let $H=F\setminus\{\ell\}$. Let $g$ be the part of this row of $B+C$ on source coordinates without the point index $a$. Then
		\begin{equation}\label{eq:lower-line-cancellation}
		 gz=\sqrt p\,\bigl(z_H-\bar r^{|H|/2}|\varnothing\rangle\bigr)^*
		       +\epsilon_\varnothing\langle\varnothing|,
		 \qquad
		 \epsilon_\varnothing=\sqrt p\,\bar r^{(|F|-1)/2}-\sqrt r.
		\end{equation}
		The first term has norm at most $\sqrt p$ and is supported on nonempty subsets of $H$. For a query to $\ell$, we have $g(I-\Pi_\ell^-)=(gz)z^*$. For a query to $u\in\ell\setminus\{a\}$, let $F_u=N(u)\setminus(N(S)\cup N(a))$. Grouping the output line indices in $F_u$ gives
		\begin{equation}\label{eq:lower-on-line-point}
		 g\Pi_u^-=\sqrt p\,(gz)\otimes|z_{F_u}\rangle
		                   m_{\{\ell\}\cup F_u}^*.
		\end{equation}
		Thus this point query multiplies the norm of each term in \eqref{eq:lower-line-cancellation} by $\sqrt p$.
	\end{lemma}
	\begin{proof}
		The subset term of $B$ and the line removal term $C$ give
		\[
		 g=\sqrt p\,z_H^*\otimes z^*+\langle\varnothing|\otimes\langle\{\ell\}|.
		\]
		Multiplying in the $\ell$ coordinate and using $z^*z=1$ and $\langle\{\ell\}|z=-\sqrt r$ gives
		\[
		 gz=\sqrt p\,z_H^*-\sqrt r\,\langle\varnothing|.
		\]
		The empty subset coordinate of $z_H$ is $\bar r^{|H|/2}$. Separating this coordinate gives exactly \eqref{eq:lower-line-cancellation}, since $|H|=|F|-1$. Deleting one coordinate from the unit vector $z_H$ leaves a vector of norm at most one, so the nonempty part has norm at most $\sqrt p$. The line query formula follows from $I-\Pi_\ell^-=zz^*$.

		For a point $u\in\ell\setminus\{a\}$, the positive bit is fixed to zero, so the corresponding difference uses the negative point projection. The only line through both $a$ and $u$ is $\ell$. The free lines in the negative point query are therefore $\{\ell\}\cup F_u$. Its part without the point index is $\sqrt p\,z\otimes z_{F_u}$. Multiplying its $\ell$ coordinate by $g$ gives the same vector $gz$, now with the additional factor $\sqrt p$, and proves \eqref{eq:lower-on-line-point}. The two remaining vectors have norm one, and the query leaves the indices in $H$ unchanged.
	\end{proof}

	The remaining estimates count how the source indices restrict the planted edge. For example, a line in the source line set and the queried line meet in at most one point, which fixes the planted point if it exists. The incidence structure therefore limits how many output blocks can use a given source coordinate. Lemma~\ref{lem:lower-counting} converts these counts into a bound valid for arbitrary vectors.

	\begin{lemma}[A bound for $E_v$]\label{lem:lower-query-errors}
		For every vertex $v$ and vector $f$ in the negative basis,
		\[
		 \|E_vf\|^2
		 =O(1)\sum_{S,T}\bigl(1+qp^2(|S|+|T|)\bigr)^2|f_{S,T}|^2.
		\]
	\end{lemma}
	\begin{proof}
		Fix a source $(S,T)$, and write $s=|S|$ and $t=|T|$. We first bound the endpoint terms by partial matchings. For the remaining terms, we identify a point or line in the source that restricts the planted edge, count the possible edges, and multiply this count by the squared local norm. The empty subset contribution has no line to use in this way; its small coefficient is instead controlled by the points in $S$.

		\paragraph{Contributions at the planted endpoints.}
		For a queried point $u$, let $C_u$ be the restriction of $C$ to edges with planted point $u$. A target has no line index through $u$, and $C_u$ removes only the planted line. Thus every contributing source $(S,T)$ contains exactly one line through $u$. That line determines the planted edge and the target label. Conversely, each target reads one source with coefficient one. The map therefore copies distinct source coordinates into distinct target coordinates, so it is a partial matching matrix and $\|C_u\|\le1$. Planting preserves $x_u$, so its contribution is zero. Since the positive bit at $u$ is one, the remaining contribution to $E_u$ is $C_u(I-\Pi_u^-)$, of norm at most one.

		For a queried line $\ell$, a target has no point index on $\ell$. The point removal map therefore uses a source $(S,T)$ with exactly one point in $S$ on $\ell$. This point determines the planted edge and the target label. As above, distinct contributing sources have distinct targets. The map is a partial matching multiplied by $\sqrt{\bar p}$, so its norm is at most one. Since $S$ contains a point on $\ell$, its basis function vanishes whenever $x_\ell=1$. The negative line projection is consequently zero on this source, and the contribution to $E_\ell$ is the map itself.

		\paragraph{Contributions from point indices.}
		For a query to $u\in\ell\setminus\{a\}$, consider the point removal term. Every source it uses contains the planted point $a$. Choosing $a\in S$ leaves at most one possible planted line, namely the line through $a$ and $u$. There are therefore at most $s$ compatible planted edges. For each edge, group the free line indices through $u$. Point removal has coefficient $\sqrt{\bar p}$, while the part of the point query vector without the index $u$ has norm $\sqrt p$. The squared local norm is at most $p\bar p\le p$. Lemma~\ref{lem:lower-counting} therefore assigns each source squared weight at most $ps\le qp^2s$.

		The empty part of \eqref{eq:lower-line-cancellation} is also controlled by point indices. For a local block, let $c=q-|F|$ be the number of lines through $a$ covered by its row's point set. This set is contained in every source read by the term, so \eqref{eq:lower-covered-lines} gives $c\le s$. Using $p\bar r^{q-1}=r$,
		\[
		 \epsilon_\varnothing
		 =\sqrt p\,\bar r^{(|F|-1)/2}(1-\bar r^{c/2})
		 \le\sqrt p\,rc\le\sqrt p\,rs.
		\]
		The bound follows from $|F|\ge1$ and the union bound $1-\bar r^{c/2}\le c(1-\sqrt{\bar r})\le cr$. In particular, $\epsilon_\varnothing=0$ when $c=0$.

		For a queried line $\ell$, the planted point can be any of its $q$ points. The squared coefficient is at most $pr^2s^2$, so the total squared weight is at most $qpr^2s^2$. For a queried point $u\in\ell\setminus\{a\}$, choose one of the $q$ lines through $u$ and one of its other $q-1$ points as $a$. Equation~\eqref{eq:lower-on-line-point} supplies an extra squared factor $p$, giving total squared weight at most $q(q-1)p^2r^2s^2$. Within each edge, the local blocks have orthogonal targets and disjoint source supports; distinct edges also have orthogonal targets. Since $qp\ge1$, Lemma~\ref{lem:lower-counting} bounds both totals by
		\[
		 O(q^2p^2r^2s^2)\le O\bigl((qp^2s)^2\bigr),
		\]
		where we used $r\le p$.

		\paragraph{Contributions from line indices.}
		For the nonempty part of a query to $\ell$, the source line set $T$ contains a line $h\in H$. This line differs from $\ell$ and passes through the planted point $a$. For each choice of $h\in T$, its intersection with the queried line fixes $a$, if an intersection exists. Thus at most $t$ planted edges can use this source. The local norm in \eqref{eq:lower-line-cancellation} is at most $\sqrt p$, so each edge contributes at most $p$ to the squared bound. Lemma~\ref{lem:lower-counting} gives total squared weight at most $pt$.

		For the nonempty part of a query to $u\in\ell\setminus\{a\}$, the source still contains a line in $H$: none of these lines passes through $u$, so the point query leaves their indices unchanged. Choose such a line from $T$, then choose $a$ among its $q$ points. The pair $a,u$ determines at most one line $\ell$, giving at most $qt$ compatible planted edges. Equation~\eqref{eq:lower-on-line-point} gives local norm at most $p$, whose square is $p^2$. The total squared weight is therefore at most $qp^2t$.

		For a queried line $h\in N(a)\setminus\{\ell\}$, planting preserves $x_h$, so its contribution to $E_h$ is zero. The positive bit is fixed to zero. In the line removal term, the $h$ coordinate is absent, and $\|\langle\varnothing|\Pi_h^-\|=\sqrt r$. Every source used by this term still contains $\ell$, since the query only changes the $h$ coordinate. Choosing $\ell\in T$ fixes $a$ as the intersection of $\ell$ and $h$, if it exists. There are at most $t$ compatible edges, each with squared local norm at most $r$, giving total squared weight at most $rt$.

		For a point query $u\notin\ell$, planting preserves the queried bit. It remains to consider $C$. When $u$ is added to $S$, the newly covered lines through $a$ form $(N(u)\setminus N(S))\cap N(a)$. This set contains at most one line, since $u\ne a$ and two distinct points lie on at most one line. If no new line is covered, the local query formulas agree before and after line removal, so the contribution to $E_u$ is zero. Otherwise, let $h$ be that new line. It is fixed to zero in the positive space but is free in the part of the negative block without the point index $u$.

		The correction $C$ removes only $\ell$. Since the target cannot contain the index $h$, the source must already omit $h$. On the subset coordinates this is the functional $\langle\varnothing|_h$, which selects the coordinate without $h$. On the point coordinate, $h$ is already excluded by the presence of $u$, so the map is the identity. After omitting unaffected coordinates, the local map is
		\[
		 (I\otimes\langle\varnothing|_h)\oplus1.
		\]
		For comparison, $z_h$ represents the normalized function $\mathbf1[x_h=0]/\sqrt{\bar r}$ by \eqref{eq:lower-zero-lines}. Taking its inner product with a line function gives $\sqrt{\bar r}$ times that function's value at $x_h=0$. Thus $z_h^*$ accounts for fixing the line bit to zero, using both line coordinates. We compare the actual map with $(I\otimes z_h^*)\oplus1$, which satisfies the following identity for the point projections:
		\[
		 \Pi_u^+\bigl((I\otimes z_h^*)\oplus1\bigr)
		 =\bigl((I\otimes z_h^*)\oplus1\bigr)\Pi_u^-.
		\]
		Indeed, it sends $\sqrt p(z_{F_u}\otimes z_h)\oplus\sqrt{\bar p}$ to $\sqrt pz_{F_u}\oplus\sqrt{\bar p}$, and its adjoint sends the latter vector back. Here $F_u$ consists of the other free lines through $u$. The difference between the two maps is zero on the part with the point index. On the part without that index it has norm
		$\|\langle\varnothing|-z_h^*\|=\sqrt{2(1-\sqrt{\bar r})}=O(\sqrt r)$, and either point projection restricted to this part has norm $\sqrt p$. Subtracting the displayed identity expresses the contribution as two terms, each bounded by the product of these norms. Its local norm is therefore $O(\sqrt{pr})$. Every source still contains $\ell$, since $u\notin\ell$ and the query at $u$ leaves that line index unchanged. There are at most $t$ choices of $\ell\in T$ and $q$ choices of $a\in\ell$. Multiplying these $qt$ possibilities by the squared local bound $O(pr)$ gives $O(qprt)\le O(qp^2t)$.

		These cases cover every possible queried vertex. For a point query, the point is $a$, another point on $\ell$, or a point outside $\ell$, all treated above. For a line query, the line is $\ell$, another line through $a$, or a line outside $N(a)$. In the last case planting preserves the bit, and removing the index $\ell$ leaves the queried line's local projection unchanged. Both contributions are therefore zero. The same reasoning accounts for the zero contributions already omitted in the other cases, by Lemma~\ref{lem:lower-planting} and the local query formulas.

		A compatible planted edge may be counted more than once if several source lines determine it; this only increases the upper bound. The endpoint contractions contribute at most $O(\|f\|^2)$. Combining this with the counts assigns each source coordinate total squared weight
		\[
		 O\bigl(1+ps+(qp^2s)^2+(p+qp^2)t\bigr)
		 =O\bigl((1+qp^2(s+t))^2\bigr).
		\]
		Here \eqref{eq:lower-probabilities} gives $p\le qp^2$. Lemma~\ref{lem:lower-counting} now gives the claimed bound.
	\end{proof}

	\paragraph{Applying the column weight.}

	The squared counting bound grows quadratically with the number of indices, while the squared column weight decreases at the matching rate. Their product is uniformly bounded, which gives the required bound on $E_vD$.
	Since $kqp^2\le2$, we have $d_{S,T}(1+qp^2(|S|+|T|))\le2$. Applying Lemma~\ref{lem:lower-query-errors} to $Df$ therefore gives $\|E_vDf\|=O(\|f\|)$, and hence $\|E_vD\|=O(1)$.

	\Needspace{14\baselineskip}
	\begin{proof}[Proof of Theorem~\ref{thm:incidence-lower-bound}]
		By \eqref{eq:lower-error-split} and Lemmas~\ref{lem:lower-numerator}, \ref{lem:lower-query-errors}, and~\ref{lem:lower-line-count},
		\begin{align*}
		 \max_v\frac{\|\Gamma\circ\Delta_v\|}{\|\Gamma\|}
		 &=O\!\left(\frac1{q\sqrt{\log q}}+\frac{\sqrt p(1+qr)}{k}\right)\\
		 &\le O\!\left(\frac1{q\sqrt{\log q}}+qp^{5/2}(1+qr)\right)
		 =O\!\left(\frac1{q\sqrt{\log q}}\right).
		\end{align*}
		The last bound uses $p=q^{-7/8+o(1)}$ and $qr=\ln q/8$. Applying the standard adversary bound of Lemma~\ref{lem:lower-adversary} proves $Q_2(\operatorname{gc}_{G_q})=\Omega(q\sqrt{\log q})$.

		For arbitrary sufficiently large $n$, choose a power of two $q$ with $2q^2\le n<8q^2$ and add isolated vertices to $G_q$. Restricting their input bits to zero recovers the original problem, so the lower bound is $\Omega(\sqrt{n\log n})$.
	\end{proof}

    \section*{Acknowledgments}

    AC is supported by ERC Starting Grant 101163189. ASG received support from the National Science Foundation (grant 26-17356). SP acknowledges support from the Dutch Ministry of Education, Culture, and Science through Gravitation project ``Challenges in Cyber Security - 024.006.037'' for this work. XZ is supported by NSERC Grant RGPIN-2024-06493.

    \paragraph{Use of generative AI.}
    OpenAI's GPT-6 Astra proposed the core proof ideas of this paper and was used extensively in developing the results and preparing the manuscript. For the upper bounds, it first produced an algorithm and analysis with query exponent $13/20$. By studying the ideas in that algorithm, we obtained, with further assistance from GPT-6 Astra, the simpler algorithm with exponent $92/139$ presented here. In parallel, continued work with GPT-6 Astra led to the improved exponent $11/17$.

    For the lower bound, we built on Belovs's approach~\cite{belovs2026tight}, with substantial assistance from GPT-6 Astra in developing the additional constructions and estimates yielding the $\Omega(\sqrt{n\log n})$ bound.

    The authors retain full responsibility for all mathematical claims and the final text.

    \begingroup
    \small
    \raggedright
    \AddToHookNext{cmd/thebibliography/after}{\addcontentsline{toc}{section}{References}}
    \bibliography{references}
    \endgroup

    \newpage
    
	\appendix
    \phantomsection
    \addcontentsline{toc}{section}{Appendix}
    \addtocontents{toc}{\protect\setcounter{tocdepth}{-1}}
	\section{Standard quantum procedures}\label{sec:quantum-proofs}

    \begin{restatedlemma}{lem:quantum}
Let a known set of $a$ labels contain $r$ marked vertices. With $\Ot(\sqrt a)$ queries, quantum counting returns an integer $\widehat r\in[0,a]$ that, except with the assigned error probability, is zero exactly when $r=0$ and otherwise satisfies
        \[
        r/2\le\widehat r\le2r,
        \qquad |\widehat r-r|=O(\sqrt r).
        \]
        Given $\widehat r>0$, a batch of $s$ independent uniform marked samples uses a fixed query limit $\Ot((s+1)\sqrt{a/\widehat r})$, reporting failure if insufficient samples are obtained. On a good estimate, sampling succeeds except with the assigned error probability, and the uniform state over marked labels can be prepared to inverse polynomial precision in $\Ot(\sqrt{a/\widehat r})$ queries.
\end{restatedlemma}
    \begin{proof}
		The claims follow from quantum counting and amplitude amplification~\cite{BHMT}. Count to additive $\sqrt r/10+1/100$ and round. For sampling, use verified randomized Grover trials of length $O(\sqrt{a/\widehat r})$, allowing $O(s+\log(1/\epsilon))$ trials. On a good estimate each trial succeeds with constant probability and, conditional on success, returns a uniform marked label. Independent trials give the stated samples and failure bound. For coherent preparation, use amplitude amplification with the lower bound $\widehat r/(2a)$ on the marked fraction~\cite{YLC}. Its query limit is fixed by the estimate and precision.
	\end{proof}

    \begin{restatedlemma}{lem:or}
Suppose polynomially many tests have known polynomially bounded query limits $q_i$. Each accepts every positive input for its own edge list and rejects negative inputs satisfying its stated bound. If all these bounds hold on a global negative input, their OR can be tested in $\Ot(\sqrt{\sum_iq_i^2})$ queries, accepting every positive global input.
\end{restatedlemma}
    \begin{proof}
		Apply standard variable-time search with known costs~\cite{VT}, with error reduction. The restriction to negative inputs causes no difficulty: in a group of $r$ comparable costs, a uniform coherent choice of child accepts with probability at least $(1-\delta)/r$ if any child is positive, regardless of the other children. On an allowed negative input its acceptance probability is at most $\delta$. Reducing each child's error to $\delta\le 1/(4r)$ costs an $O(\log r)$ factor and separates the two cases by a constant factor. Standard amplification distinguishes these cases; grouping costs between consecutive powers of two gives the stated bound.
	\end{proof}

    \section{Collision tests}\label{sec:tests}
    The tests below have fixed query limits. Their bounds are required on negative inputs; a collision is accepted except with the assigned error probability whether or not those bounds hold.

    \begin{algorithm}[H]
		\caption{\covertest}\label{alg:cover}
		\small
		\textbf{Input:} $(G,U,b)$, where $U$ is a known vertex cover.
		\begin{steps}
			\item Group the positive degree vertices of $U$ by $d\le\deg_G(u)<2d$.
			\item \textbf{for each} nonempty group, with $d$ a power of two:
			\item \ind Estimate its marked count by $\widehat r$.
			\item \ind If $\widehat r d>2b$, return \Yes; if $\widehat r=0$, skip this group.
			\item \ind Amplify the following trial $O(\sqrt{\widehat r})$ times:
			\item \ind\ind Prepare a uniform marked group vertex and search its neighborhood.
			\item \ind Return \Yes\ if a verified marked edge is found.
			\item \textbf{end for}
			\item \textbf{return} \No.
		\end{steps}
	\end{algorithm}
	
	\begin{restatedlemma}{lem:cover}
Let $U$ be a known vertex cover of $G$. The procedure $\covertest(G,U,b)$ uses $\Ot(\sqrt{|U|}+\sqrt b)$ queries. It accepts every positive input and rejects negative inputs satisfying $\vol(G,U,x)\le b$.
\end{restatedlemma}
	\begin{proof}
		Failed preparations in the amplified trial are nonaccepting; the trial accepts only a verified marked edge. For a group of size $a$ with $r$ marked vertices, a good estimate satisfies $r/2\le\widehat r\le2r$. On an allowed negative input, $rd\le b$, so the count check cannot accept. If a collision has a marked endpoint in the group, a trial finds a verified edge with probability $\Omega(1/r)$. Every collision has an endpoint in $U$, proving acceptance on positive inputs.
		
		Counting costs $\Ot(\sqrt a)$. Each trial before amplification costs $\Ot(\sqrt{a/\widehat r}+\sqrt d)$, so the total for a group is $\Ot(\sqrt a+\sqrt{\widehat r d})\le\Ot(\sqrt a+\sqrt b)$. Summing over the logarithmically many groups proves the bound. For $b=0$, any nonzero estimate accepts immediately; an empty cover needs no queries.
	\end{proof}

	\begin{algorithm}[H]
		\caption{\bipartitetest}\label{alg:bipartite}
		\small
		\textbf{Input:} $(G,A,B,k)$, with $|A|=a$ and $|B|=b$.
		\begin{steps}
			\item If a side is empty, return \No; otherwise let $k\gets\min(k,b)$.
			\item Estimate the number of marked vertices in $B$ by $\widehat r$.
			\item If $\widehat r=0$, return \No; if $\widehat r>2k$, return \Yes.
			\item Let $s=\min\{\widehat r,\max\{1,\lceil(a\widehat r^2/b)^{1/3}\rceil\}\}$.
			\item Run quantum walk search on ordered lists of $s$ marked vertices from $B$.
			\item \ind \textbf{Setup:} prepare the uniform superposition over lists, allowing repetition.
			\item \ind \textbf{Update:} replace a random entry by a fresh uniform marked vertex.
			\item \ind \textbf{Check:} search $A$ for a marked neighbor of some list entry.
			\item Use gap bound $1/s$ and accepting fraction bound $s/(3\widehat r)$.
			\item \textbf{return} the outcome of the walk search.
		\end{steps}
	\end{algorithm}
	
	\begin{restatedlemma}{lem:bipartite}
Let $G$ be bipartite with sides $A,B$ of sizes $a,b$.

        The procedure $\bipartitetest(G,A,B,k)$ accepts every positive input and rejects negative inputs satisfying $|M\cap B|\le k$, with query complexity
        \[
        \Ot\bigl(\sqrt a+\sqrt b+(a^2bk)^{1/6}\bigr).
        \]
\end{restatedlemma}
	\begin{proof}
		Condition first on a good count estimate. If $r=0$, no collision exists. If $\widehat r>2k$, returning \Yes\ is safe because this cannot occur on a negative input satisfying $r\le k$. Otherwise $r/2\le\widehat r\le2r$ and $1\le s\le\widehat r$.
		
		The walk has uniform stationary distribution on ordered lists from $(M\cap B)^s$. Refreshing one of $s$ independent coordinates gives spectral gap $1/s$ when $r\ge2$; when $r=1$, there is only one list and its check suffices. If a collision has endpoint $v\in B$, a list containing $v$ is accepting. Its stationary probability is at least
		\[
		1-(1-1/r)^s\ge1-e^{-s/r}\ge\frac{s}{r+s}
		\ge\frac{s}{3\widehat r}.
		\]
		On a negative input no list is accepting. Since the graph is known, adjacency to the stored list is computed without queries. An update prepares a uniform index $i$ and a uniform marked vertex $w$, and writes the list $y$ obtained by replacing $x_i$ with $w$. If $y\ne x$, then $i$ and $w$ are determined by $x$ and $y$ and can be uncomputed. If $y=x$, then $w=x_i$ can be uncomputed, and the index register, which is uniform, is returned to its initial state. Lemma~\ref{lem:quantum} supplies coherent preparation and updates, giving
		\[
		S=\Ot\left(s\sqrt{b/\widehat r}\right),\qquad
		U=\Ot\left(\sqrt{b/\widehat r}\right),\qquad
		C=\Ot(\sqrt a).
		\]
		Lemma~\ref{lem:walk} therefore bounds the cost, including counting, by
		\[
		\Ot\left(s\sqrt{\frac b{\widehat r}}+\sqrt b
		+\sqrt{\frac{a\widehat r}{s}}\right).
		\]
		When $(a\widehat r^2/b)^{1/3}$ lies between $1$ and $\widehat r$, the first and last terms balance at $(a^2b\widehat r)^{1/6}$, up to a constant from rounding. If it is below $1$, these terms are $O(\sqrt b)$; if it exceeds $\widehat r$, they are $O(\sqrt a)$. Since every estimate reaching the walk satisfies $\widehat r\le2k$, the fixed query limits give \eqref{eq:bipartite} on every execution. Good estimates and the walk guarantee acceptance on every positive input, and rejection on every negative input satisfying the count bound. Counting errors are included in the assigned error probability.

		For the implementation errors, preparing a state within Euclidean distance $\eta$ of its target gives, by preparation, phase reversal, and uncomputation, a reflection within operator norm $4\eta$ of the exact reflection. The same estimate holds for preparations controlled by a stored list. Reducing the checking error probability to $O(\eta^2)$ gives a coherent check with operator norm error $O(\eta)$ on the subspace where its work registers are initialized. There are only polynomially many such operations, so choosing $\eta$ sufficiently small and inverse polynomial in $n$ makes the sum of these errors at most $\epsilon$. A telescoping expansion of the products of operations then bounds the change in the final state by $\epsilon$. The additional precision costs only logarithmic factors.
	\end{proof}
	
	\begin{restatedlemma}{lem:few}
Let $G$ have at most $n$ vertices and let $U$ be a known vertex cover. There is a test using $\Ot\bigl(\sqrt n(1+k)^{1/6}\bigr)$ queries that accepts every positive input and rejects negative inputs satisfying $|M\cap U|\le k$.
\end{restatedlemma}
	\begin{proof}
		Make a bipartite graph whose first side is a copy of all vertices and whose second side is a copy of $U$, joined by the original incidences. Apply the bipartite test of Lemma~\ref{lem:bipartite}. Each query to a copied input bit costs one original query, and a marked edge exists in the copy exactly when it exists in the original graph.
	\end{proof}

	\Needspace{20\baselineskip}
	\section{Constructing the edge partition}\label{sec:partition-proof}

    \begin{lemma}[Marked counts in a piece]\label{lem:count}
		Let a piece have sides $A,B$, sizes $a,b$, and $m$ edges. Its degrees on $A$ lie between $m/(\gamma a)$ and $\gamma m/a$, and similarly on $B$. If its marked volume is at most $\ell$, let $p=\min\{1,\gamma\ell/m\}$. Then $|M\cap A|\le ap$ and $|M\cap B|\le bp$. These marked count bounds remain valid after edge deletions when the side labels are fixed.
	\end{lemma}
	\begin{proof}
		On either side the maximum and minimum degrees differ by at most $\gamma$. Comparing them with the average degree gives the stated degree bounds for pieces from Lemma~\ref{lem:partition}. Each marked vertex on $A$ contributes at least $m/(\gamma a)$ to marked volume. There are therefore at most $\gamma a\ell/m$ such vertices, and at most $a$ trivially. The argument on $B$ is identical. Edge deletions do not change the input bits of the fixed labels, so the marked count bounds remain valid.
	\end{proof}
    
	The improved algorithm from \Cref{sec:algorithm} needs pieces with comparable degrees on each side and disjoint edge lists. The following construction (adapted from \cite[Lemma~11]{CDR19}) uses only the known graph. It supplies the degree bounds used in Lemma~\ref{lem:count}; its query cost is zero. Algorithm~\ref{alg:partition} gives the construction.
	
	\begin{algorithm}[H]
		\caption{\partitionedges}\label{alg:partition}
		\small
		\textbf{Input:} A bipartite graph $G$.
		\begin{steps}
			\item Initialize an empty list of pieces.
			\item \textbf{while} edges remain:
			\item \ind Group side $i$ by degrees in $(d_i/2,d_i]$, for powers of two $d_i$.
			\item \ind Choose a pair of groups containing the most edges; copy those edges to $F$.
			\item \ind Repeatedly remove from $F$ a vertex on side $i$ of degree below $d_i/\gamma$.
			\item \ind Record $F$ and its maximum degree on each side.
			\item \ind Delete exactly $E(F)$ from $G$.
			\item \textbf{end while}
			\item \textbf{return} the recorded pieces.
		\end{steps}
	\end{algorithm}
	
	\begin{lemma}[Exact partition]\label{lem:partition}
		Algorithm~\ref{alg:partition} partitions a known bipartite graph into $j=O((\log n)^3)$ nonempty pieces, without input queries. It returns triples $(F,d_1,d_2)$, where $F$ is an exact edge list on its incident labels and every vertex on side $i$ has degree in $[d_i/\gamma,d_i]$, for $\gamma=32(\log n)^2$. Marked volume is additive over the pieces.
	\end{lemma}
    \begin{proof}
		Edges removed from $F$ during trimming remain in $G$ for subsequent iterations.
		
		If the current graph has $m$ edges, some pair contains at least $m/(4(\log n)^2)$ edges. Each selected side has at most $2m/d_i$ vertices. Charging a trimmed edge to its first removed endpoint gives total charge at most
		\[
		\frac{2m}{d_1}\frac{d_1}{\gamma}
		+\frac{2m}{d_2}\frac{d_2}{\gamma}
		=\frac{m}{8(\log n)^2}.
		\]
		At least $m/(8(\log n)^2)$ edges are recorded. Repeated removal of this fraction gives $j=O((\log n)^3)$. Each piece has the claimed degree bounds. Recording the actual maximum degree on each side preserves the degree ratio and ensures that the recorded bounds do not exceed the maximum degree of the input graph. Its edge list is disjoint from all others, proving both exactness and additivity.
	\end{proof}

\end{document}